\documentclass[a4paper,10pt]{article}
\usepackage{stmaryrd}
\usepackage{amsfonts}
\usepackage{bbm}
\usepackage{amscd}
\usepackage{mathrsfs}
\usepackage{latexsym,amssymb,amsmath,amscd,amscd,amsthm,amsxtra}
\usepackage[utf8]{inputenc}
\usepackage[T1]{fontenc}
\usepackage{lmodern}
\usepackage{amssymb}
\usepackage{nicefrac,mathtools,enumitem}
\usepackage{microtype}
\usepackage{lipsum}
\usepackage{mathtools}

\usepackage{tikz}
\usepackage[none]{hyphenat}

\newtheorem{thm}{Theorem}[section]
\newtheorem{lem}[thm]{Lemma}
\newtheorem{cor}[thm]{Corollary}
\newtheorem{pro}[thm]{Proposition}
\newtheorem{ex}[thm]{Example}
\newtheorem{rmk}[thm]{Remark}
\newtheorem{defi}[thm]{Definition}

\newcommand{\be }{\begin{equation}}
\newcommand{\ee }{\end{equation}}

\newcommand{\pf}{\noindent{\bf Proof.}\ }

\def\qed{\hfill ~\vrule height6pt width6pt depth0pt}

\newcommand{\br}[1]{   [ \cdot,    \cdot  ]   }

\newcommand{\g}{\mathfrak g}

\newcommand{\Ad}{\mathrm{Ad}}

\newcommand{\ad}{\mathrm{ad}}

\newcommand {\sfPhi}{\mathsf{\Phi}}

\newcommand {\IR}{\mathbb{R}}

\newcommand{\spr}{/\!\!/}

\begin{document}
\title{\small\bf
IRREGULAR RIEMANN-HILBERT CORRESPONDENCE,
ALEKSEEV-MEINRENKEN DYNAMICAL r-MATRICES AND DRINFELD TWISTS}

\author{\small XIAOMENG XU}

\date{}
\newcommand{\Addresses}{{
  \bigskip
  \footnotesize

  \textsc{DEPARTMENT OF MATHEMATICS, MASSACHUSETTS INSTITUTE OF TECHNOLOGY, CAMBRIDGE, MA 02139, USA}\par\nopagebreak
  \textit{E-mail address}: \texttt{xxu@mit.edu}

}}


\maketitle

\begin{abstract}
In 2004, a new approach to the Ginzburg-Weinstein linearization theorem for a quasitriangular Lie
bialgebra $(\g,r)$ was suggested by Enriquez, Etingof and Marshall. This approach is based on solving a system of PDEs for a gauge transformation between the classical r-matrix $r$ and the Alekseev-Meinrenken dynamical r-matrix.
In this paper, we explain that preferred gauge transformations can be constructed as connection maps for a certain irregular Riemann-Hilbert problem (provided $r$ is the standard classical r-matrix). Our construction is based on earlier works by Boalch. Along the way, we give a symplectic geometric interpretation of the PDEs, as a symplectic neighborhood version of the Ginzburg-Weinstein linearization theorem. We then prove that for a semisimple Lie algebra $\g$, any solution of the PDEs for the gauge transformation is the semiclassical limit of an admissible Drinfeld twist. As a byproduct, we find a surprising relation between the connection maps and Drinfeld twists.
\end{abstract}

\setcounter{tocdepth}{1}
\tableofcontents
\section{Introduction and main results}
In the study of non-commutative Weil algebra \cite{AM}, Alekseev and Meinrenken introduced a particular dynamical $r$-matrix $r_{\scriptscriptstyle \rm AM}$, which is an important special case of classical dynamical $r$-matrices (\cite{Felder}, \cite{EV}). Let $\g$ be a complex reductive Lie algebra and $t\in S^2(\g)^2$ the element corresponding to a nondegenerate bilinear form on $\g$, then $r_{\scriptscriptstyle \rm AM}$, as a map from $\g^*$ to $\g\wedge\g$, is defined by
\begin{eqnarray*}
r_{\scriptscriptstyle \rm AM}(x):=({\rm id}\otimes \phi({\rm ad}_{x^{\vee}}))(t), \ \forall x\in\frak g^*,
\end{eqnarray*}
where $x^{\vee}=(x\otimes \rm{id})(t)$ and $\phi(z):=-\frac{1}{z}+\frac{1}{2}\rm{cotanh}\frac{z}{2},$ $z\in \mathbb{C}\setminus 2\pi i\mathbb{Z}^*$.
Remarkably, this r-matrix came to light naturally in two different applications, i.e., in the context of equivariant cohomology \cite{AM} and in the description of a Poisson structure on the chiral WZNW phase space compatible with classical $G$-symmetry \cite{Feher}.

Let $r\in \g\otimes\g$ be a classical $r$-matrix such that $r+r^{2,1}=t$ (thus $(\frak g,r)$ is a quasitriangular Lie bialgebra). In \cite{EEM}, Enriquez, Etingof and Marshall constructed formal Poisson isomorphisms between the formal Poisson manifolds $\frak g^*$ and $G^*$ (the dual Poisson Lie group). Here $\frak g^*$ is equipped with its Kostant-Kirillov-Souriau structure, and $G^*$ with its Poisson Lie structure given by $r$.
Their result relies on constructing a formal map $g:\frak g^*\rightarrow G$ satisfying the following gauge transformation equation (as identity of formal maps $\g^*\rightarrow \wedge^2(\g)$)
\begin{eqnarray}\label{introequation}
g_1^{-1}d_2(g_1)-g_2^{-1}d_1(g_2)+(\otimes^2{\rm Ad}_g)^{-1}r_0+\langle {\rm id}\otimes {\rm id}\otimes x,[g_1^{-1}d_3(g_1),g_2^{-1}d_3(g_2)]\rangle=r_{\scriptscriptstyle \rm AM},
\end{eqnarray}
Here $r_0:=\frac{1}{2}(r-r^{2,1})$, $g_1^{-1}d_2(g)=\sum_i g^{-1}\frac{\partial g}{\partial \xi^{i}}\otimes e_i$ is viewed as a formal
function $\frak g^*\rightarrow \frak g^{\otimes 2}$, $\{e_i\}$ is a basis of $\frak g$, $\{\xi^i\}$ the corresponding coordinates on $\frak g^*$ and $g_i^{-1}d_j(g_i) = (g_1^{-1}d_2(g_1))^{i,j}$.

Two constructions of solutions of \eqref{introequation} are given: the first one uses the theory of the classical Yang-Baxter equation and gauge transformations; the second one relies on the theory of quantization of Lie bialgebras.
The result in \cite{EEM} may be viewed as a generalization of the formal version of \cite{GW}, in which Ginzburg and Weinstein proved the existence of a Poisson diffeomorphism between the real Poisson manifolds $k^*$ and $K^*$, where $K$ is a compact Lie group and $k$ is its Lie algebra. Different approaches to similar results in the subject of linearization of Poisson structures can be found in \cite{Anton} and \cite{Boalch1}.

The main purpose of the present paper is to give explicit solutions of the above equation (provided $r$ is a standard classical $r$-matrix). The solutions will be constructed as the monodromy of certain differential equations with irregular singularities. This construction enables us to understand the geometric meaning of equation \eqref{introequation} and clarify its relation with irregular Riemann-Hilbert correspondence. In the following, we give an introduction of the main results.

\subsection*{Symplectic geometric construction}
Our first result is to give a symplectic geometric interpretation of equation \eqref{introequation}. The construction is as follows.

In Section \ref{Geometryconstruction}, we introduce a symplectic slice $\Sigma$ of $T^*G$ and its Poisson Lie analogue, a symplectic submanifold $\Sigma'$ of the Lu-Weinstein symplectic double $\Gamma$ (locally isomorphic to $G\times G^*$) \cite{Lu}. Associated to any $g\in {\rm Map}(\frak g^*,G)$, we define a local diffeomorphism
\begin{eqnarray}\label{Fg}
F_g:(\Sigma,\omega)\rightarrow (\Sigma',\omega').
\end{eqnarray}
Then we have
\begin{thm}\label{third}
$F_g$ is a local symplectic isomorphism from $(\Sigma,\omega)$ to $(\Sigma',\omega')$ if and only if $g\in {\rm Map}(\frak g^*,G)$ satisfies equation \eqref{introequation}.
\end{thm}
We will see that this theorem builds the bridge between the gauge equation \eqref{introequation} and a certain irregular Riemann-Hilbert correspondence. It is also the key gradient for the proof of Theorem \ref{main}. The proof of this theorem is given in the technical appendix.

\subsection*{Enriquez-Etingof-Marshall gauge transformations and Stokes phenomenon}
In Section \ref{defineC},  we construct preferred solutions of equation \eqref{introequation} via Stokes phenomenon. For this, we consider the meromorphic connection on the trivial holomorphic principal $G$-bundle $P$ over $\mathbb{P}^1$,
\begin{eqnarray}\label{nabla}
\nabla=d-(\frac{A_0}{z^2}+\frac{1}{2\pi i}\frac{x}{z})dz
\end{eqnarray}
where $A_0,x\in \frak g$. We assume that $A_0\in \frak t_{\rm reg}$ and once fixed, the only variable is $x\in\g\cong\g^*$ (identification via a nondegenerate bilinear form on $\g$).
Then we consider the monodromy of $\nabla$ from $0$ to $\infty$, known as the connection matrix $C(x)$ of $\nabla$, which is the ratio of two canonical solutions of $\nabla F=0$, one of which is around $\infty$ and another is on one chosen Stokes sector at $0$. Varying $x\in\g^*$, we obtain the {\em connection map} $C:\g^*\rightarrow G;\ x\mapsto C(x)$. More precisely, $C$ is only defined on an open dense subset of $\g^*$, i.e., for those $x$ such that $\nabla$ is non–resonant. See Section \ref{defineC} for more details. The main result of this paper is
\begin{thm}\label{main}
For any $A_0\in\frak t_{\rm reg}$, the connection map $C\in {\rm Map}(\frak g^*,G)$ is a solution of equation \eqref{introequation}.
\end{thm}
The background and reformulation of this theorem is given in Section 3, and its proof is given in Section \ref{irregularmap}.

Let $G^*$ be the dual Poisson Lie group associated to the standard r-matrix. In \cite{EEM}, Enriquez, Etingof and Marshall associate a (formal) Poisson isomorphism $S_g: \g^*\rightarrow G^*$ to any solution $g$ of \eqref{introequation}.  In Section \ref{ss:Stokes PL}, we show that the Poisson map $S_C$, associated to the connection map $C$ for any $A_0\in\frak t_{\rm reg}$, coincides with the irregular Riemann-Hilbert map $\nu:\g^*\rightarrow G^*$ relating $x\in\frak g^*$ to the Stokes matrices of $\nabla$. In particular, it recovers the remarkable result due to Boalch which states that for any $A_0\in\frak t_{\rm reg}$, the irregular Riemann-Hilbert map $\nu:\frak g^*\rightarrow G^*$ is a local analytic Poisson isomorphism. In other words, we have the following commutative diagram:

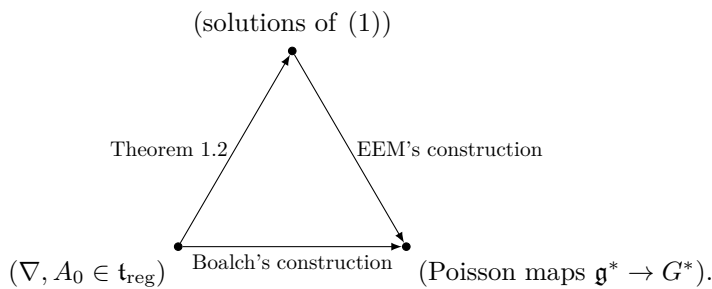
\[\begin{tikzpicture}[>=latex,mydot/.style={draw,circle,inner
    sep=1pt},every label/.style={scale=1},scale=1]

  \foreach \i in {0}{
  \node[mydot, fill=black, label=240:$(\nabla{,}\, A_0\in\frak t_{\rm reg})$]                at (-1.5+9*\i, -2.595)    (p\i0) {};
  \node[mydot,fill=black,label=90:(solutions of \eqref{introequation})]      at (+9*\i,0) (p\i1) {};
  \node[mydot,fill=black,label=-60:(Poisson maps $\g^*\rightarrow G^*$).]     at (1.5+9*\i,-2.595)    (p\i2) {};
}
\begin{scope}[<-]
    \draw (p01)--node[left,scale=.8]{Theorem \ref{main}} (p00);
    \draw (p02)--node[below,scale=.8]{Boalch's construction} (p00);
    \draw (p02)--node[right,scale=.8]{EEM's construction} (p01);
\end{scope}
\end{tikzpicture}\]
We also develop a symplectc neighborhood version of Ginzburg-Weinstein linearization. See Section \ref{ss:Stokes PL} for more details.


In Section \ref{irregularmap}, we give a proof of Theorem \ref{main} and clarify the relation between the gauge equation \eqref{introequation} and a certain irregular Riemann-Hilbert correspondence. This is motivated by and based on Boalch's works, e.g.  \cite{Boalch2} \cite{BoalchG} \cite{Boalch3}, on the study of the geometry of moduli spaces of meromorphic connections on a trivial holomorphic principal $G$-bundle on Riemann surfaces with divisors.

\subsection*{Drinfeld twists and connection maps}
Having proved that the connection map satisfies the gauge transformation equation \eqref{introequation}, in Section \ref{twist} we further discuss its relation with Drinfeld twist. This is based on and motivated by a series of works of Enriquez and Etingof (among others) on the theory of quasi-Hopf algebras and dynamical twist quantization.

Let $(U(\g),m,\Delta,\varepsilon)$ denote the universal enveloping algebra of $\g$ with the product $m$, the coproduct $\Delta$ and the counit $\varepsilon$. Let $U(\g)\llbracket\hbar\rrbracket$ be the corresponding topologically free $\mathbb{C}\llbracket\hbar\rrbracket$-algebra.
In \cite{EEM}, the equation \eqref{introequation} was interpreted as the classical limit of a vertex-IRF transformation equation \cite{EN} between a dynamical twist $J_d(x)\in {\rm Map}(\g^*,U(\g)^{\hat{\otimes}2}\left\llbracket\hbar\right\rrbracket)$ and a constant twist $J_c\in U(\g)^{\hat{\otimes}2}\left\llbracket\hbar\right\rrbracket$. Here $J_d(x)$ and $J_c$ are respectively the twist quantization of $r_{\scriptscriptstyle \rm AM}$ and $r$ associated to an admissible associator $\Phi$ \cite{EH}. As a result, the semiclassical limit of a vertex-IRF transformation
$\rho\in {\rm Map}(\g^*,U(\g)\left\llbracket\hbar\right\rrbracket)$, as a formal map from $\g^*$ to the formal group $G$, gives rise to a formal solution of \eqref{introequation}.

According to \cite{EE}\cite{EEM}, an admissible Drinfeld twist $J\in U(\g)^{\hat{\otimes} 2}\llbracket\hbar\rrbracket$ (killing the associator $\Phi$) produces such a vertex-IRF transformation.
Thus, in particular, the semiclassical limit of the twist $J$ provides a solution of \eqref{introequation}. We then study the gauge actions on the space of solutions of \eqref{introequation} and on the space of admissible Drinfeld twists. This study enables us to show that
\begin{thm}
For the case of a semisimple Lie algebra $\g$, any formal solution $g\in{\rm Map}_0(\g^*,G)$ of \eqref{introequation} is the semiclassical limit of an admissible twist.
\end{thm}
The proof of this theorem is given in Section \ref{sec43}.
As a consequence, in Section \ref{sec44} we introduce
\begin{cor} For any $A_0\in\frak t_{\rm reg}$ and the associated connection map $C\in {\rm Map}(\g^*,G)$, there exists a Drinfeld twist killing the associator $\Phi$ whose semiclassical limit is $C$.
\end{cor}

In particular, let $\Phi$ be the Knizhnik-Zamolodchikov (KZ) associator $\Phi_{KZ}$, which is the monodromy from $1$ to $\infty$ of the KZ equation on $\mathbb{P}^1$ with three simple poles at $0$, $1$, $\infty$. Naively, the confluence of two simple poles at $0$ and $1$ in the KZ equation leads to a degree two pole, while the monodromy representing KZ associator becomes the connection matrix $C_{\hbar}$ for an irregular Riemann-Hilbert problem. Then Theorem \ref{Cequation} and the above corollary indicate that the monodromy $C_{\hbar}$ may give a certain Drinfeld twist killing $\Phi_{KZ}$. Indeed, an explicit construction of the Drinfeld twist is given by Toledano Laredo \cite{TL}. See \cite{TLXu} for a further discussion and a relation to the present paper.

\vspace{5mm}

The organisation of this paper is as follows. The next section gives the background material and a symplectic geometric interpretation of the equation \eqref{introequation}. Section 3 defines the connection map $C:\frak g^*\rightarrow G$ of meromorphic connections $\nabla$ and states that $C$ gives rise to a solution of \eqref{introequation}. Section 4 gives the background material on the moduli space of meromorphic connections over surfaces and irregular Riemann-Hilbert correspondence. In the second part of Section 4, we give a proof of Theorem \ref{main} by studying one explicit case of the correspondence in details. Section 5 discusses the quantum version of equation \eqref{introequation}, i.e., the vertex-IRF transformation equation, and relates any solution of \eqref{introequation} to the semiclassical limit of an admissible Drinfeld twist. In particular, it formulates a surprising relation between connection matrices and Drinfeld twists. The Appendix studies the Poisson submanifolds of the Lu-Weinstein symplectic double and gives a proof of Theorem \ref{third}.

\subsection*{Acknowledgements}
\noindent
I give my warmest thanks to Anton Alekseev for his encouragements as well as inspiring discussions and insightful suggestions. The proof of Theorem \ref{main} is based on Philip Boalch's work, to whom I am very grateful for sharing valuable ideas. My sincere thanks are also extended to Benjamin Enriquez, Pavel Etingof, Giovanni Felder, Marco Gualtieri, Anna Lachowska, David Li-Bland, Ian Marshall, Nicolai Reshetikhin, Pavol $\rm{\check{S}}$evera, Valerio Toledano Laredo, Alan Weinstein, Ping Xu and Chenchang Zhu for their useful discussions, suggestions and interests in this paper. This work is partially supported by the Swiss National Science Foundation (SNSF) grants P2GEP2-165118, NCCR SwissMAP and the European Research Council (ERC) project MODFLAT.

\section{Gauge transformations of $r$-matrices and symplectic geometry}\label{Geometryconstruction}
In Section \ref{sec21}, we recall the Enriquez-Etingof-Marshall gauge transformation \cite{EEM} between a classical r-matrix and the Alekseev-Meinrenken dynamical r-matrix \cite{AM}. In Section \ref{geoconstruction}, we introduce a symplectic submanifold $\Sigma'$ of Lu-Weinstein symplectic double \cite{Lu}. Then in Theorem \ref{equivalent}, we use $\Sigma'$ to give a symplectic geometric interpretation of the gauge transformation equation.
\subsection{Dynamical r-matrices and gauge transformations}\label{sec21}
Let $\frak g$ be a complex reductive Lie algebra, $t\in S^2(\g)^{\g}$ the element corresponding to a nondegenerate invariant bilinear form on $\g$.
First recall that an element $r\in\frak g\otimes \frak g$ is a classical $r$-matrix if its symmetric part $r+r^{2,1}\in S^2(\frak g)^{\frak g}$ and it satisfies the classical Yang-Baxter equation:
\begin{eqnarray*}
[r^{1,2},r^{1,3}]+[r^{1,2},r^{2,3}]+[r^{1,3},r^{2,3}]=0.
\end{eqnarray*}
Throughout this paper, we will denote by $r_0:=\frac{1}{2}(r-r^{2,1})$ the skew-symmetric part of a classical $r$-matrix $r$.

A dynamical analog of a classical $r$-matrix is as follows.
\begin{defi}
A classical dynamical $r$-matrix, with respect to a Lie subalgebra $\eta\subset \frak g$, is an $\eta$-equivariant map $r:\eta^*\rightarrow \frak g\otimes\frak g$ such that $r+r^{2,1}\in S^2(\frak g)^{\frak g}$ and $r$ satisfies the dynamical Yang-Baxter equation (CDYBE):
\begin{eqnarray}\label{CDYBE}
{\rm Alt}(dr)+[r^{1,2},r^{1,3}]+[r^{1,2},r^{2,3}]+[r^{1,3},r^{2,3}]=0,
\end{eqnarray}
where ${\rm Alt}(dr(x))\in \wedge^3\frak g$ is the skew-symmetrization of $dr(x)\in \eta\otimes\frak g\otimes\frak g\subset \frak g\otimes\frak g\otimes\frak g$ for all $x\in\eta^*$.
\end{defi}
{\bf Alekseev-Meinrenken dynamical r-matrix.} In the distinguished special case $\eta=\frak g$, the Alekseev-Meinrenken dynamical $r$-matrix $r_{\scriptscriptstyle \rm AM}: \g^*\rightarrow \g\otimes\g$ is defined by
\begin{eqnarray*}
r_{\scriptscriptstyle \rm AM}(x)=({\rm id}\otimes \phi({\rm ad}_{x^{\vee}}))(t), \ \forall x\in\frak g^*,
\end{eqnarray*}
where $x^{\vee}=(x\otimes {\rm id})(t)$ and $\phi(z):=-\frac{1}{z}+\frac{1}{2}{\rm cotanh}\frac{z}{2},$ $z\in \mathbb{C}\setminus 2\pi i\mathbb{Z}^*$. Taking the Taylor expansion of $\phi$ at $0$, we see that $\phi(z)=\frac{z}{12}+O(z^2)$, thus $\phi({\rm ad}_x)$ is well-defined (the maximal domain of definition of $\phi({\rm ad}_x)$ contains all $x\in \frak g^*$ for which the eigenvalues of ${\rm ad}_x$ lie in $\mathbb{C}\setminus 2\pi i \mathbb{Z}^*$).
One can check that $r_{\scriptscriptstyle \rm AM}+\frac{t}{2}$ is a classical dynamical $r$-matrix.

{\bf Enriquez-Etingof-Marshall gauge transformation equation.} Denote by $G$ the formal group with Lie algebra $\g$ and by ${\rm Map}_0(\g^*,G)$ the space of formal maps $g:\g^*\rightarrow G$ such that $g(0)=1$, i.e., the space of maps of the form $e^{u}$, where $u\in\g\otimes \hat{S}(\g)_{\ge 0}$ ($\hat{S}(\g)$ is the degree completion of the symmetric algebra $S(\g)$).
The following theorem states the existence of formal solutions of equation \eqref{introequation}.
\begin{thm}{\rm (\cite{EEM})}\label{EEM}
Let $r$ be a classical $r$-matrix with $r+r^{2,1}=t$ and $r_0:=\frac{1}{2}(r-r^{2,1})$. Then there exists a formal map $g\in {\rm Map}_0(\frak g^*,G)$, such that
\begin{eqnarray}\label{gaugeequation}
g_1^{-1}d_2(g_1)-g_2^{-1}d_1(g_2)+(\otimes^2{\rm Ad}_g)^{-1}r_0+\langle {\rm id}\otimes {\rm id}\otimes x,[g_1^{-1}d_3(g_1),g_2^{-1}d_3(g_2)]\rangle=r_{\scriptscriptstyle \rm AM},
\end{eqnarray}
Here $g_1^{-1}d_2(g)(x):=\sum_i g^{-1}\frac{\partial g}{\partial{\xi^{i}}}(x)\otimes e_i$ is viewed as a formal
function $\frak g^*\rightarrow \frak g^{\otimes 2}$, $\{e_i\}$ is a basis of $\frak g$, $\{\xi^i\}$ the corresponding coordinates on $\frak g^*$ and $g_i^{-1}d_j(g_i) = (g_1^{-1}d_2(g_1))^{i,j}$.
\end{thm}
We will call \eqref{gaugeequation} the gauge transformation equation, and denote its left hand side by $r_0^g\in \rm{Map}(\frak g^*,\frak g\wedge\frak g)$.
In \cite{EEM}, this equation is proven to be the classical limit of vertex-IRF transformation between dynamical twists (see section \ref{twist}) and the authors give two constructions of the formal solutions of equation \eqref{gaugeequation} based on a formal calculation and a quantization of Lie bialgebras respectively.
In the following, we will give a geometric interpretation and construct explicit solutions of equation \eqref{gaugeequation} via Stokes phenomenon. Instead of the formal setting, we will work on a local theory.

\subsection{Symplectic geometric construction}\label{geoconstruction}
{\bf The symplectic manifold $(\Sigma,\omega)$.} We follow the convention from last section. Let $\frak t\subset \frak g$ be a maximal abelian subalgebra and $\frak t'$ the complement of the affine root hyperplanes: $\frak t':= \{\Lambda\in \frak t~|~ \alpha(\Lambda)\notin 2\pi i\mathbb{Z}\}$. In the following, $\frak t'$ is regarded as a subspace of $\g^*$ via the isomorphism $\g\cong\g^*$ induced by the nondegenerate bilinear form.

Let $\Sigma$ be a cross-section of $T^*G\cong G\times \g^*$ (identification via left multiplication), defined by
\begin{eqnarray*}
\Sigma:=\{(h,\lambda)\in G\times \frak g^*~|~\lambda \in \frak t'\}.
\end{eqnarray*}
One can check that $\Sigma$ is a symplectic submanifold of $T^*G$ with the canonical symplectic structure (see \cite{GS} Theorem 26.7). The induced symplectic structure $\omega$ on $\Sigma$ is given for any tangents $v_1=(X_1,R_1),v_2=(X_2,R_2)\in \frak g\times \frak t^*$ at $(h,\lambda)$ by
\begin{eqnarray}\label{omegasigma}
\omega(v_1,v_2)=\langle R_1,X_2\rangle-\langle R_2,X_1\rangle+\langle \lambda,[X_1,X_2]\rangle.
\end{eqnarray}
\\
{\bf The Poisson-Lie analogue of $(\Sigma,\omega)$.}
Let $r\in\g\otimes\g$ be a classical $r$-matrix with $r+r^{2,1}=t$. Let $G^*$ be the simply connected dual Poisson Lie group associated to the quasitriangular Lie biaglebra $(\frak g, r)$ and $D$ the double Lie group with Lie algebra $\frak d=\frak g\Join\frak g^*$ which is locally diffeomorphic to $G\times G^*$ (see e.g \cite{Lu1}). A natural symplectic structure on $D$ is given by the following bivector,
\begin{eqnarray*}
\pi_D=\frac{1}{2}(r_d\pi_0+l_d\pi_0),
\end{eqnarray*}
where $\pi_0\in \frak d\wedge \frak d$ such that $\pi_0(\xi_1+X_1,\xi_2+X_2)=\langle X_1,\xi_2\rangle-\langle X_2,\xi_1\rangle$ for $\xi_1+X_1,\xi_2+X_2\in \frak d^*\cong\frak g^*\oplus \frak g$.

Following \cite{Lu}, the Lu-Weinstein double symplectic groupoid is the set
\begin{eqnarray*}
\Gamma:=\{(h,h^*,u,u^*)~|~h,u\in G,h^*,u^*\in G^*, hh^*=u^*u\in D\}
\end{eqnarray*}
with a unique Poisson structure $\pi_{\Gamma}$ such that the local diffeomorphism $(\Gamma,\pi_{\Gamma})\rightarrow (D,\pi_D)$: $(h,h^*,u,u^*)\mapsto hh^*$ is Poisson. We define a submanifold $\Sigma'$ of $\Gamma$, as a Poisson Lie analogue of $\Sigma$, by
\begin{eqnarray*}
\Sigma':=\{(h,h^*,u,u^*)\in \Gamma~|~h^*\in e^{\frak t'}\subset G^*\}.
\end{eqnarray*}
Here $e$ denotes the exponential map with respect to the Lie algebra $g^*$.
In the Appendix, we will prove that $\Sigma'$ is a symplectic submanifold of $(\Gamma,\pi_{\Gamma})$. Now let us take this fact and denote the induced symplectic structure on $\Sigma'$ by $\omega'$.
Note that the map
\begin{eqnarray*}
\Sigma'\rightarrow \Sigma; \ (h,e^{\lambda},u,u^*)\mapsto (h,\lambda)
\end{eqnarray*}
expresses $\Sigma'$ as a cover of a dense subset of $\Sigma$. Thus $\Sigma$ and $\Sigma'$ are locally diffeomorphic to each other.
\\
\\
{\bf Symplectic maps between $(\Sigma,\omega)$ and $(\Sigma',\omega')$.} Let ${\rm Map}_0(\g^*,G)$ be the space of maps $g:\g^*\rightarrow G$ such that $g(0)=1$. Associated to any $g\in {\rm Map}_0(\frak g^*,G)$, we define a map $F_g:\Sigma\rightarrow \Sigma'$ by
\begin{eqnarray}\label{DiffFg}
F_g(h,\lambda):=(g({\rm Ad}_h\lambda)h,e^{\lambda},u,u^*), \ \forall (h,\lambda)\in\Sigma,
\end{eqnarray}
where $u\in G, u^*\in G^*$ are determined by the identity $g({\rm Ad}_h\lambda)he^{\lambda}=u^*u$ (understood to hold in the double Lie group $D$). Note that $F_g$ is well-defined for the elements $(h,\lambda)\in \Sigma$ sufficiently near $(1,0)\in G\times\frak t^*$. This is because for these $(h,\lambda)$, $g({\rm Ad}_h\lambda)he^\lambda$ in the double Lie group $D$ is sufficiently near the unit, thus $g({\rm Ad}_h\lambda)he^{\lambda}=u^*u$ uniquely determines $u$ and $u'$. So we can think of $F_g$ defined on a local chart and this is enough for our purpose.
\begin{thm}\label{equivalent}
$F_g$ is a local symplectic isomorphism from $(\Sigma,\omega)$ to $(\Sigma',\omega')$ if and only if $g\in {\rm Map}_0(\frak g^*,G)$ satisfies the gauge transformation equation \eqref{gaugeequation}, $r_0^{g}=r_{\scriptscriptstyle \rm AM}$.
\end{thm}
\pf See the Appendix.
\\
\\
{\bf The case of a standard classical $r$-matrix.} Let $T\subset G$ be a maximal torus with Lie algebra $\frak t\subset \frak g$. Let $B_{\pm}$ denote a pair of opposite Borel subgroups with $B_+\cap B_-=T$. We take the standard $r$-matrix given by
\begin{eqnarray}\label{rmatrix}
r:=\frac{1}{2}t+\frac{1}{2}\sum_{\alpha\in \sfPhi_+}E_{\alpha}\wedge E_{-\alpha},
\end{eqnarray}
where $\sfPhi_+$ is the positive root system corresponding to the Borel subgroup $B_+$.
In this case, the simply connected dual Poisson Lie group associated to $(\frak g,r)$ is
\begin{eqnarray*}
G^*=\{(b_-,b_+,\Lambda)\in B_-\times B_+\times \frak t~|~\delta(b_-)\delta(b_+)=1,\delta(b_+)=\rm{exp}(\pi i\Lambda)\},
\end{eqnarray*}
where $\delta:B_\pm\rightarrow T$ is the group homomorphism corresponding to the projection $\frak g\rightarrow \frak t$.
Thus $\Sigma'$ is a submanifold of the double
\begin{eqnarray*}
\Gamma:\{h,(b_-,b_+,\Lambda), u,(c_-,c_+,\Lambda_c)~|~hb_{\pm}=c_{\pm}u\}\subset (G\times G^*)^2,
\end{eqnarray*}
defined by
\begin{eqnarray*}
\Sigma':=\{(h,(e^{-\pi i \Lambda},e^{\pi i \Lambda},\Lambda),u,(c_+,c_-,\Lambda_c))\in\Gamma~|~he^{\pm \pi i \Lambda}=c_{\pm}u, \Lambda\in \frak t'\},
\end{eqnarray*}
where $\frak t'\subset \frak t$ is the complement of the affine root hyperplanes.

\section{Enriquez-Etingof-Marshall gauge transformations via Stokes phenomenon}\label{defineC}
In Section \ref{sec31}, we introduce a meromorphic connection $\nabla$ on a G-bundle over $\mathbb{P}^1$ and its canonical solutions. The presentation and notation are close to those in \cite{Boalch1}. In Section \ref{connectiondata}, we introduce the connection map of $\nabla$, and then state our main result, Theorem \ref{Cequation}, which says that the connection map gives rise to a solution of equation \eqref{gaugeequation}. In Section \ref{ss:Stokes PL}, we explain how Theorem \ref{StokesPoissonLie} from Boalch's work \cite{Boalch1} on Stokes matrices and Poisson Lie groups can be recovered from Theorem \ref{Cequation}.

\subsection{Meromorphic connections and canonical solutions}\label{sec31}
Let $G$ be a complex reductive Lie group, $T\subset G$ a maximal
torus, and $\frak t\subset\g$ the Lie algebra of $T$.
Let $\sfPhi\subset\frak t^*$ be the corresponding root system of $\g$,
and $\frak t_{\rm reg}$ the set
of regular elements in $\frak t$.

Let $P$ be the holomorphically trivial principal $G$-bundle on $\mathbb{P}^1$. We consider the meromorphic connection on $P$ of the form
\begin{eqnarray*}
\nabla:=d-(\frac{A_0}{z^2}+\frac{1}{2\pi i}\frac{x}{z})dz,
\end{eqnarray*}
where $A_0,x\in\g$. We assume henceforth that $A_0\in\frak t_{\rm reg}$ and once it is fixed, the only variable is $x\in \frak g\cong \frak g^*$ (via the nondegenerate bilinear form on $\g$). Note that the connection $\nabla$ has an order 2 pole at origin and (if $x\neq 0$) a first order pole at $\infty$.

\begin{defi}
The {\it Stokes rays} of the connection $\nabla$ are the rays $\IR_{>0}
\cdot\alpha(A_0)\subset\mathbb{C}^*$, $\alpha\in\sfPhi$. The {\it Stokes sectors}
are the open regions of $\mathbb{C}^*$ bounded by two adjacent Stokes rays.
\end{defi}

Let us choose an initial sector ${\rm Sect}_0$ at $0$ and a branch of ${\rm log}(z)$ on ${\rm Sect}_0$.
Then we label the Stokes rays $d_1, d_2,...,d_{2l}$ going in a positive sense and starting on the positive edge of ${\rm Sect}_0$. Set ${\rm Sect}_i={\rm Sect}(d_i,d_{i+1})$ for the open sector bounded by the rays $d_i$ to $d_{i+1}$. (Indices are taken modulo $2l$, so ${\rm Sect}_0={\rm Sect}(d_{2l}, d_1)$).

To each sector ${\rm Sect}_i$, there is a canonical solution $F_i$ of $\nabla F=0$ with prescribed asymptotics
in the $i$-th supersector $\widehat{{\rm Sect}_i}:={\rm Sect}(d_i-\frac{\pi}{2},d_i+\frac{\pi}{2})$.
In particular, the following result is proved in \cite{BJL} for $G=GL
_n(\mathbb{C})$, in \cite{BoalchG} for $G$ reductive, and in \cite{BTL1} for an
arbitrary affine algebraic group.
Denote by $\delta(x)$ the projection of $x$ onto $\frak t$ corresponding to
the root space decomposition $\g=\frak t\bigoplus_{\alpha\in\sfPhi}\g_
\alpha$.

\begin{thm}\label{jurk}
On each sector ${\rm Sect}_i$, there
is a unique holomorphic function $H_i:{\rm Sect}_i\to G$ such that the function
\[F_i=H_i\cdot e^{-\frac{A_0}{z}}\cdot z^{\frac{\delta(x)}{2\pi i}}\]
satisfies $\nabla F_i=0$, and $H_i$ can be analytically continued to $\widehat{{\rm Sect}_i}$ and then $H_i$ is asymptotic to $1$ within $\widehat{{\rm Sect}_i}$.
\end{thm}


\subsection{Connection maps and Enriquez-Etingof-Marshall gauge transformations}\label{connectiondata}
Following the convention in \cite{BTL1}, we say that the meromorphic connection $\nabla=d-(\frac{A_0}{z^2}+\frac{1}{2\pi i}\frac{x}{z})dz$ is {\it non--resonant}
at $z=\infty$ if the eigenvalues of $\frac{1}{2\pi i}{\rm ad}(x)$ are not positive integers. The following fact is well-known (see e.g \cite{Wasow} for $G={\rm GL}_n(\mathbb{C})$).

\begin{lem}\label{le:nr dkz}
If $\nabla$ is non--resonant, there is a unique holomorphic function
$H_\infty:\mathbb{P}^1\setminus\{0\}\to G$ such that $H_\infty(\infty)=1$, and the function $F_\infty=H_\infty
\cdot z^{\frac{x}{2\pi i}}$ is a solution of $\nabla F=0$.
\end{lem}

Now let us consider the two solutions of $\nabla F=0$:
\begin{eqnarray*}
&&F_0 \ on \ \rm{
Sect}_0,\\
&&F_\infty \ on \ a \ neighbourhood \ of \ \infty \ slit \ along \ d_1,
\end{eqnarray*}
We define the {\em connection
matrix} $C(x)\in G$ (with respect to the chosen ${\rm Sect}_0$) by
\[F_\infty=F_0\cdot C(x).\]
Here $F_\infty$ is extended along a path in ${\rm  Sect}_0$ then the identity is understood to hold in the domain of definition of $F_0$.

Varying $x\in\g^*_{\rm nr}$, we obtain the {\em connection map} $C:\g^*_{\rm nr}\to G$.
Here $\g^*_{\rm nr}\subset\g^*$ is the dense open set corresponding to the set of elements $x$
such that the eigenvalues of $\frac{1}{2\pi i}\ad(x)$ do not contain positive integers (provided we identify $\g\cong\g^*$). Note that the connection map $C$ depends on the choice of $A_0$, and some discrete data (the initial Stokes sector and the branch of ${\rm log}(z)$).

The Stokes sector ${\rm Sect}_0$ determines a partition of the root system
$\sfPhi$ of $\g$ as follows. Let $\Pi_+$ and $\Pi_-$ be the sets of Stokes
rays which one crosses when going from ${\rm Sect}_0$ to the opposite sector ${\rm Sect}_l$ in the
counterclockwise and clockwise directions respectively. Then $\sfPhi=\sfPhi_+
\sqcup\sfPhi_-$, where
\[\sfPhi_\pm=\{\alpha\in\sfPhi|\,\alpha(A_0)\in\ell, \ell\in\Pi_\pm\}=-\sfPhi_\mp.\]

Now let us consider the equation \eqref{gaugeequation}, in which $r_0$ is the skew-symmetric part of the standard $r$-matrix associated to the positive root system $\sfPhi_+$.
Our main theorem states that
\begin{thm}\label{Cequation}
The connection map $C\in {\rm Map}(\frak g^*_{\rm nr},G)$ is a solution of the gauge transformation equation \eqref{gaugeequation}.
\end{thm}
A proof will be given at the end of Section \ref{irregularmap}. The basic idea is to verify its symplectic geometric counterpart, i.e., to prove $F_{C}:(\Sigma,\omega)\rightarrow (\Sigma',\omega')$ is a symplectic map.

\subsection{Stokes matrices and linearization of $G^*$}\label{ss:Stokes PL}

Given the initial Stokes sector ${\rm Sect}_0$ and the determination of $\log(z)$
with a cut along the Stokes ray $d_1$, the Stokes matrices are the transition matrices between the canonical solutions $F_0$ on ${\rm Sect}_0$ and $F_l$ on the opposite sector ${\rm Sect}_l$, when they are continued along the two possible paths in the punctured disk joining these sectors.
Thus the {\it Stokes matrices} of $\nabla=d-(\frac{A_0}{z^2}+\frac{1}{2\pi i}\frac{x}{z})dz$ with respect to
to ${\rm Sect}_0$ are the elements $S_\pm(x)$ of $G$ determined by
\[F_0=F_{l}\cdot S_+(x) e^{\delta(x)}, \ \ \ \ \
F_{l}=F_0\cdot S_-(x)
\]
where $\delta(x)$ takes the projection of $\g$ onto $\frak t$, and the first (resp. second) identity is understood to hold in ${\rm Sect}_l$
(resp. ${\rm Sect}_0$) after $ F_0$ (resp. $ F_{l}$)
has been analytically continued counterclockwise.
The connection matrix $C(x)$ is related to the Stokes matrices $S_\pm(x)$
by the following monodromy relation (from the fact that a simple positive loop around $0$ is a simple negative loop around $\infty$).

\begin{lem}\label{le:monodromy reln}
The following holds
\[ C(x)^{-1} e^{\delta(x)} C(x)=S_-(x) S_+(x) e^{\delta(x)}\]
\end{lem}

\begin{rmk}
In \cite{Boalch1}, Boalch proves that the connection map $C$ is a Duistermaat twist \cite{Duistermaat} by using the monodromy relation. 
\end{rmk}

Recall that the Stokes sector ${\rm Sect}_0$ determines a partition of the root system
$\sfPhi=\sfPhi_+
\sqcup\sfPhi_-$. Let $U_\pm\subset G$
be the unipotent subgroups with Lie algebra $\frak u_\pm=\bigoplus_
{\alpha\in\sfPhi_\pm}\g_\alpha$, and $B_\pm$ the corresponding opposite Borel subgroups.
It follows from \cite{BoalchG} that the Stokes matrices
$S_+(x),S_-(x)$ lie in $U_+,U_-$ respectively. Varying $x\in\g^*$, we therefore obtain the Stokes map
\[\nu:\g^*\to G^*; \ x\mapsto (e^{-\frac{\delta (x)}{2}} S_-(x)^{-1},
e^{-\frac{\delta(x)}{2}} S_+(x)e^{\delta(x)},\frac{\delta(x)}{2\pi i}).\]
Here $G^*$ is the dual Poisson Lie group defined in Section \ref{Geometryconstruction}, \begin{eqnarray*}
G^*=\{(b_-,b_+,\Lambda)\in B_-\times B_+\times \frak t~|~\delta(b_-)\delta(b_+)=1,\delta(b_+)=\rm{exp}(\pi i\Lambda)\}.
\end{eqnarray*}
The relation between the Stokes map $\nu$ and the theory of Poisson Lie groups can be shown as follows.

It follows from Theorem \ref{equivalent} and Theorem \ref{Cequation} that the map
\[F_{C}:(\Sigma,\omega)\rightarrow (\Sigma',\omega'); \ (h,\lambda)\mapsto (C({\rm Ad}_h\lambda)h,e^{\lambda},u,u^*)\] is a local symplectic isomorphism. This map is equivariant with respect to the symplectic $T$-actions on $\Sigma$ and $\Sigma'$, which are respectively given by
\begin{eqnarray*}
a\cdot (h,\lambda)=(ha,\lambda), \ \ \ \ \ \ \ a\cdot (h,e^{\lambda},u,u^*)=(ha, e^{\lambda},u,u^*), \ \forall  a\in T.
\end{eqnarray*}
Define two maps $P:\Sigma\rightarrow \frak g^*$, $P':\Sigma'\rightarrow G^*$ whose fibres are the $T$ orbits
\begin{eqnarray*}
P(h,\lambda)={\rm Ad}^*_h\lambda, \ \forall (h,\lambda)\in \Sigma, \ \ \ \ \ P'(h,e^{\lambda},u,u^*)=d_he^{\lambda}, \ \forall (h,e^{\lambda},u,u^*)\in \Sigma'.
\end{eqnarray*}
Here $d$ denotes the left dressing transformation of $G$ on $G^*$.
By using the monodromy relation \eqref{le:monodromy reln}, we see that the Stokes map $\nu:\frak g^*\rightarrow G^*$ is the unique map such that the following diagram commutes:
$$
\begin{CD}
\Sigma @> F_{C}>> \Sigma' \\
@V P_1 VV @V P_2 VV \\
\frak g^* @> S_C >> G^*
\end{CD}
$$
One can check that $P_1$ and $P_2$ are Poisson maps (see e.g the Appendix), where $\g^*$ is equipped with its standard Kirillov--Kostant--Souriau
Poisson structure and $G^*$ the dual Poisson Lie group structure. Therefore, the $T$-reduction of the symplectic isomorphism $F_{C}$ gives rise to the following remarkable result due to Boalch.
\begin{thm}{\rm\cite{Boalch1}}\label{StokesPoissonLie}
For any $A_0\in\frak t_{\rm reg}$, the Stokes map $\nu:\frak g^*\rightarrow G^*$ is a local analytic Poisson isomorphism.
\end{thm}
The above discussion justifies that $F_C:(\Sigma,\omega)\rightarrow (\Sigma',\omega')$ is a symplectic neighborhood version of Ginzburg-Weinstein linearization.

\section{Irregular Riemann-Hilbert correspondence}\label{irregularmap}
In Section \ref{sec51} and \ref{sec52}, we recall respectively the symplectic moduli spaces of meromorphic connections on a trivial holomorphic principal $G$-bundle and the symplectic spaces of monodromy/Stokes data. We mainly follow the papers \cite{Boalch2}\cite{BoalchG}\cite{Boalch3} of Boalch, in which these symplectic spaces are found and described both explicitly and from an infinite dimensional viewpoint (generalising the Atiyah-Bott approach). Then in Section \ref{sec53}, we discuss the irregular Riemann-Hilbert correspondence and analyze in details the Riemann-Hilbert map $\nu$ for the meromorphic connections with one simple pole and one order two pole. In Section \ref{GaugeviaRH} Proposition \ref{RHmap}, we show that $\nu$ can be explicitly expressed by the connection map defined in Section 3.2. It enables us to finally give a proof of Theorem \ref{Cequation} by using the symplectic nature of the Riemann-Hilbert map $\nu$.

\subsection{Moduli spaces of meromorphic connections and the spaces of monodromy data}\label{sec51}
Let $D=\sum_{i=1}^m k_i(a_i)>0$ be an effective divisor on $\mathbb{P}^1$ and $P$ a holomorphically trivial principal $G$-bundle. Let us consider the meromorphic connections on $P$ with poles on $D$. They can be described explicitly as follows.
Let $z$ be a local coordinate on $\mathbb{P}^1$ vanishing at $a_i$. Then in terms of a local trivialisation of $P$, any meromorphic connection $\nabla$ on $P$ takes the form of $\nabla=d-A$, where
\begin{eqnarray*}
A=\frac{A_{k_i}}{z^{k_i}}dz+\cdot\cdot\cdot \frac{A_1}{z}dz+A_0dz+\cdot\cdot\cdot,
\end{eqnarray*}
and $A_j\in \g$, $j\le k_i$.
\begin{defi}
A compatible framing at $a_i$ of a trivial principal $G$-bundle $P$ with a generic connection $\nabla$ is an isomorphism \ $\mathclap{^i}g_0:P_{a_i}\rightarrow G$ between the fibre $P_{a_i}$ and $G$ such that the leading coefficient of $\nabla$ is inside $\frak t_{\rm reg}$ in any local trivialisation of $P$ extending $g_0$.
\end{defi}

Let us choose, at each point $a_i$, an irregular type \[\mathclap{^i}{A}^0:=\mathclap{^i}{A}_{k_i}\frac{dz}{z^{k_i}}+\cdot\cdot\cdot+\mathclap{^i}{A}_{2}\frac{dz}{z^2},\]
where $\mathclap{^i}A_{k_i}\in \frak t_{\rm reg}$ and $\mathclap{^i}A_j\in\frak t$ for $j<k_i$.
Let $\nabla=d-A$ in some local trivialisation (thus a compatible framing is an element in $G$) and $z_i$ a local coordinate vanishing at $a_i$. Then we say $(\nabla,P)$ with compatible framing \ $\mathclap{^i}g_0$ at $a_i$ has irregular type \ $\mathclap{^i}{A}^0$ if there is some formal bundle automorphism $g\in G\llbracket z_i\rrbracket$ with $g(a_i)=\mathclap{^i}g_0$, such that $$g[A]:=gAg^{-1}+dg\cdot g^{-1}=\mathclap{^i}{A}^0+\frac{\mathclap{^i}\Lambda}{z_i}dz_i$$ for some \ $\mathclap{^i}\Lambda\in\frak t$.
We denote by ${\bf a}$ the choice of the effective divisor $D=\sum_{i=1}^m k_i(a_i)$ and the chosen irregular types \ $\mathclap{^i}A^0$.
\begin{defi}{\rm (\cite{Boalch2})}
The extended moduli space $\widetilde{\mathcal{M}^*}({\bf a})$ is the set of isomorphism classes of triples $(P,\nabla,{\bf g})$, consisting of a generic connection $\nabla$ with poles on $D$ on a trivial holomorphic principal $G$-bundle $P$ over $\mathbb{P}^1$ with compatible framing ${\bf g}=( \ \mathclap{^1}g_0,..., \ \mathclap{^m}g_0)$, such that $(P,\nabla,{\bf g})$ has irregular type \ $\mathclap{^i}A^0$ at each $a_i$.
\end{defi}

Next let us recall (from \cite{Boalch2} Section 2) the building blocks $\widetilde{O}$ of the moduli space $\widetilde{\mathcal{M}^*}({\bf a})$.
Fix an integer $k\ge 2$. Let $G_k:=G(\mathbb{C}[z]/z^k)$ be the group of $(k-1)$-jets of bundle automorphisms, and let $\frak g_k =\rm{Lie}(G_k)$ be its Lie algebra, which contains elements of the form $X=X_0+X_1z+\cdot\cdot\cdot+X_{k-1}z^{k-1}$ with $X_i\in\frak g$. Let $B_k$ be the subgroup of $G_k$ of elements having constant term 1. The group $G_k$ is the semi-direct product $G\ltimes B_k$ (where $G$ acts on $B_k$ by conjugation). Correspondingly the Lie algebra of $G_k$ decomposes as a vector space direct sum and dualizing we have: $\frak g^*_k=\frak b^*_k\oplus \frak g^*$. Elements of $\frak g^*_k$ will be written as
\begin{eqnarray*}
A = A_k\frac{dz}{z^k}+\cdot\cdot\cdot +A_{1}\frac{dz}{z}
\end{eqnarray*}
via the pairing with $\frak g_k$ given by $\langle A,X\rangle:= {\rm Res_0}(A,X)=\sum_{j}(A_j,X_{j-1})$. In this way $\frak b^*_k$ is identified with the set of $A$ having zero residue and $\frak g^*$ with those having only a residue term (zero irregular part). Let $\pi_{{\rm res}}:\frak g^*_k\rightarrow\frak g^*$ and $\pi_{{\rm irr}}:\frak g^*_k\rightarrow \frak b^*_k$ denote the corresponding projections.

Now choose an element $A^0=A^0_k\frac{dz}{z^k}+\cdot\cdot\cdot+A^0_{2}\frac{dz}{z^2}$ of $\frak b^*_k$ with $A^0_i\in \frak t$ and $A_0^0\in\frak t_{\rm{reg}}$. Let $O_{A^0}\subset \frak b^*_k$ denote the $B_k$ coadjoint orbit containing $A^0$.
\begin{defi}\rm{(\cite{Boalch2})}
The extended orbit $\widetilde{O}\subset G\times \frak g^*_k$ associated to $O_{A^0}$ is
\begin{eqnarray*}
\widetilde{O}:=\{(g_0,A)\in G\times\frak g^*_k ~|~\pi_{irr}(g_0Ag_0^{-1})\in O_{A^0}\}
\end{eqnarray*}
where $\pi_{irr}:\frak g^*_k\rightarrow\frak b^*_k$ is the natural projection removing the residue.
\end{defi}
$\widetilde{O}$ is naturally a Hamiltonian $G$-manifold. Any tangents $v_1,v_2$ to $\widetilde{O}\in G\times\frak g^*_k$ at $(g_0,A)$ are of the form
\begin{eqnarray*}
v_i=(X_i(0),[A,X_i]+g_0^{-1}{R_i}g_0)\in \frak g\oplus \frak g^*_k
\end{eqnarray*}
for some $X_1,X_2\in \frak g_k$ and ${R_1},{R_2}\in \frak t^*$ (where $\frak g\cong T_{g_0}G$ via left multiplication), and the symplectic structure on $\widetilde{O}$ is given by
\begin{eqnarray}\label{omegaO}
\omega_{\widetilde{O}}(v_1,v_2)=\langle R_1,{\rm Ad}_{g_0}X_2\rangle-\langle R_2,{\rm Ad}_{g_0}X_1\rangle+\langle A,[X_1,X_2]\rangle.
\end{eqnarray}
\begin{pro}\rm{(\cite{Boalch2})}
The $G$ action $h\cdot (g_0,A):=(g_0h^{-1},hAh^{-1})$ on $(\widetilde{O},\omega_{\widetilde{O}})$ is Hamiltonian with moment map $\mu_G:\widetilde{O}\rightarrow \frak g^*$, $\mu(g_0,A)=\pi_{res}(A)$.
\end{pro}
In the simple pole case $k = 1$ we define
\begin{eqnarray*}
\widetilde{O}:=\{(h,x)\in G\times \frak g^* ~|~{\rm Ad}_hx\in\frak t''\}\subset G\times\frak g^*,
\end{eqnarray*}
where $\frak t'':= \{\Lambda\in \frak t~|~ \alpha(\Lambda)\notin \mathbb{Z}\}$. One can check that the map $\widetilde{O}\rightarrow \Sigma$; $(h,x)\mapsto (h, 2\pi i{\rm Ad}_hx)$ is an isomorphism of $\widetilde{O}$ to the symplectic slice $\Sigma$ defined in section \ref{defineC} (provided the symplectic structure on $\Sigma$ is divided by $2\pi i$).

The spaces $\widetilde{O}$ enable us to construct global symplectic moduli spaces of meromorphic connections on trivial $G$-bundles over $\mathbb{P}^1$ as symplectic quotients of the form $\widetilde{O}_1\times\cdot\cdot\cdot\times\widetilde{O}_m\spr G$ (the Hamiltonian reduction of the direct product of $m$ Hamiltonian $G$-spaces).

\begin{pro}{\rm (\cite{Boalch2})}
$\widetilde{\mathcal{M}^*}({\bf a})$ is isomorphic to the symplectic quotient
\begin{eqnarray*}
\widetilde{\mathcal{M}^*}({\bf a})\cong \widetilde{O}_1\times\cdot\cdot\cdot\times\widetilde{O}_m\spr G
\end{eqnarray*}
where $\widetilde{O}_i\subset G\times \frak g^*_{k_i}$ is the extended coadjoint orbit associated to the $B_{k_i}$ coadjoint orbit $O_{ \mathclap{^i}A^{0}}\subset \frak b^*_{k_i}$ containing the element \ $\mathclap{^i}A^{0}\in\frak b^*_k$ (the chosen irregular type at $a_i$).
\end{pro}

\subsection{Quasi-Hamiltonian $G$-spaces and symplectic spaces of monodromy/Stoke data.}\label{sec52}
In the following, we describe the symplectic space $\widetilde{\mathcal{M}}({\bf a})$ of monodromy/Stokes data for compatibly meromorphic connections $(V,\nabla,{\bf g})$ with irregular type ${\bf a}$. Let us start by briefly recalling the theory of quasi-Hamiltonian geometry \cite{AMM}.

Let $\theta$ (resp. $\bar{\theta}$) denote the left (resp. right) invariant $\g$-valued Cartan one-form on $G$. Let $\psi$ denote the canonical three-form of $G$, i.e., $\psi:=\frac{1}{6}\langle\theta,[\theta,\theta]\rangle$.
\begin{defi}\rm{(\cite{AMM})}
A complex manifold $M$ is a complex quasi-Hamiltonian $G$-space if it carries a $G$-action, a $G$-equivariant map $\mu:M\rightarrow G$ (where $G$ acts on itself by conjugation), and a $G$-invariant holomorphic two-form $\omega\in \Omega^2(M)$ such that

\begin{itemize}
  \item[(i)] $d\omega=\mu^*(\psi)$, where $\psi$ is the canonical three-form on $G$;

\item[(ii)] $\omega(v_X,\cdot)=\frac{1}{2}\mu^*(\theta+\bar{\theta},X)\in \Omega^1(M)$, for all $X\in\frak g$, where $v_X$ is the fundamental vector field $(v_X)_m=-\frac{d}{dt}(e^{tX}\cdot m)|_{t=0}$.

\item[(iii)] at each point $m\in M$, the kernel of $\omega$ is
\begin{eqnarray*}
{\rm ker} \omega_m=\{(v_X)_m~|~X\in\frak g \ such \ that \ hXh^{-1}=-X, \ where \ h:=\mu(m)\in G\}.
\end{eqnarray*}
\end{itemize}
\end{defi}
These axioms are motivated by the study of Hamiltonian loop group manifolds. See \cite{AMM} for more details.
\begin{ex}
Let $\mathcal{C}\subset G$ be a conjugacy class equipped with the conjugation $G$-action and the moment map $\mu:\mathcal{C}\rightarrow G$ given by the inclusion. Then $\mathcal{C}$ is a quasi-Hamiltonian $G$-space with the two-form $\omega$ defined by
\begin{eqnarray*}
\omega_h(v_X,v_Y)=\frac{1}{2}(\langle X, {\rm Ad}_hY \rangle-\langle Y, {\rm Ad}_hX\rangle ),
\end{eqnarray*}
for any $h\in \mathcal{C}$, $X,Y\in \frak g$ and $v_X,v_Y$ the fundamental vector field with respect to the conjugation $G$-action.
\end{ex}
Similar to the Hamiltonian reduction, we have the following moment map reduction in the quasi-Hamiltonian setting.
\begin{thm}{\rm (\cite{AMM})}
Let $(M,\omega)$ be a quasi-Hamiltonian $G$-space with moment map $\mu:M\rightarrow G$. If the quotient $\mu^{-1}(1)/G$ of the inverse image $\mu^{-1}(1)$ of the identity is a manifold, then the restriction of $\omega$ to $\mu^{-1}(1)$ descends to a symplectic form on the reduced space $M\spr G:=\mu^{-1}/G$.
\end{thm}

\begin{defi}{\rm (\cite{AMM})}
Let $M_1$ (resp. $M_2$) be a quasi-Hamiltonian $G$-space with moment map $\mu_1$ (resp. $\mu_2$). Their fusion product $M_1\circledast M_2$ is defined to be the quasi-Hamiltonian $G$-space $M_1\times M_2$, with the diagonal $G$ action, two-form
\begin{eqnarray*}
\widetilde{\omega}=\omega_1+\omega_2-\frac{1}{2}(\mu_1^*\theta,\mu_2^*\bar{\theta})
\end{eqnarray*}
and moment map
\begin{eqnarray*}
\widetilde{\mu}=\mu_1\cdot\mu_2: \ M\rightarrow G.
\end{eqnarray*}
\end{defi}
The quasi-Hamiltonian $G$-spaces from conjugacy classes can be seen as the building blocks of moduli spaces of flat connections on the trivial $G$-bundle on $\mathbb{P}^1$. Indeed, following \cite{AMM} let $\Sigma_m$ be a sphere with $m$ boundary components, the quasi-Hamiltonian reduction
\begin{eqnarray*}
\mathcal{C}_1\circledast\cdot\cdot\cdot \circledast \mathcal{C}_m\spr G
\end{eqnarray*}
of the fusion product of $m$ conjugacy classes $\mathcal{C}_i$ is isomorphic to the moduli space of flat connections on $\Sigma_m$ with the Atiyah-Bott symplectic form.

{\bf Symplectic spaces of monodromy data.}
Let us next describe the building blocks of the monodromy/Stokes data of meromorphic connections using quasi-Hamiltonian geometry. Let $T$ be a maximal torus of $G$ with Lie algebra $\frak t\subset \frak g$ and $B_{\pm}$ denote a pair of opposite Borel subgroups with $B_+\cap B_-=T$. Let us consider the family of complex manifolds (see \cite{Boalch3} for the geometrical origins of these spaces where their infinite-dimensional counterparts are described)
\begin{eqnarray*}
\widetilde{\mathcal{C}}:=\{(C,{\bf d},{\bf e},\lambda)\in G\times (B_-\times B_+)^{k-1}\times t ~|~\delta(d_j)^{-1}=
e^{\frac{\pi i\lambda}{k-1}}=\delta(e_j) \ for \ all \ j\},
\end{eqnarray*}
parameterised by an integer $k\geq 2$, where ${\bf d}=(d_1,...,d_{k-1})$, ${\bf e}=(e_1,...,e_{k-1})$ with $d_{even},e_{odd}\in B_+$,
$d_{odd},e_{even}\in B_-$, and $\delta :B_\pm\rightarrow T$ is the homomorphism whose kernel is the unipotent subgroup $U_{\pm}$.
\begin{pro}\rm{(\cite{Boalch3})}\label{k=k_i}
The manifold $\widetilde{\mathcal{C}}$
is a complex quasi-Hamiltonian $G$-space with action
\begin{eqnarray*}
g\cdot (C,{\bf d},{\bf e}, \lambda)=(Cg^{-1},{\bf d},{\bf e},\lambda)\in \widetilde{\mathcal{C}}, \ \ \forall g\in G,
\end{eqnarray*}
moment map
\begin{eqnarray*}
\mu:\widetilde{\mathcal{C}}\rightarrow G; \ \ (C,{\bf d},{\bf e},\lambda)\mapsto C^{-1}d_1^{-1}\cdot\cdot\cdot d_{k-1}^{-1}e_{k-1}\cdot\cdot\cdot e_1C,
\end{eqnarray*}
and two-form
\begin{eqnarray*}
\omega=\frac{1}{2}(\bar{\mathcal{D}},\bar{\mathcal{E}})+\frac{1}{2}\sum_{j=1}^{k-1}(\mathcal{D}_j,\mathcal{D}_{j-1})-(\mathcal{E}_j,\mathcal{E}_{j-1}).
\end{eqnarray*}
Here $\bar{\mathcal{D}}=D^*\bar{\theta}$, $\bar{\mathcal{E}}=E^*\bar{\theta}$, $\mathcal{D}_j=D^*_j\theta$, $\mathcal{E}_j=E^*_j\theta\in
\Omega^1(\widetilde{\mathcal{C}},\frak g)$ for maps $D_j$, $E_j:\widetilde{\mathcal{C}}\rightarrow G$ defined by $D_i(C,\mathbb{d},\mathbb{e},\Lambda)=d_i\cdot\cdot\cdot d_1C,$ $E_i=e_i\cdot\cdot\cdot e_1C,$ $D:=D_{k-1},$ $E:=E_{k-1}$, $E_0=D_0:=C$.
\end{pro}
\begin{rmk}
The extension of the Atiyah-Bott symplectic structure to the case of singular $\mathbb{C}^{\infty}$-connections given in \cite{Boalch2} leads to certain Hamiltonian loop group manifolds, and $\widetilde{\mathcal{C}}$ is the corresponding quasi-Hamiltonian space.
\end{rmk}
For instance, when $k=2$, $\widetilde{\mathcal{C}}_{k=2}\cong G\times G^*$. The moment map and two form are given by
\begin{eqnarray}\label{k=2}
\mu=C^{-1}b_-^{-1}b_+C, \ \ \ \omega=\frac{1}{2}(D^*\bar{\theta},E^*\bar{\theta})+\frac{1}{2}(D^*\theta,C^*{\theta})-\frac{1}{2}(E^*\theta,C^*\theta)
\end{eqnarray}
where $D=b_-C,E=b_+C$.

For $k = 1$, we define $\widetilde{\mathcal{C}}_{k=1}:=\{(h,(e^{-\pi i \lambda},e^{\pi i \lambda},\lambda))~|~h\in G,\lambda\in \frak t'\}$ which is a submanifold of $\widetilde{\mathcal{C}}_{k=2}\cong G\times G^*$, and thus inherits a $G$ action. The restriction of the two form and moment map \eqref{k=2} of $\widetilde{\mathcal{C}}_{k=2}$ to $\widetilde{\mathcal{C}}_{k=1}$ makes it into a quasi-Hamiltonian $G$-space.

Given an effective divisor $D=\sum_{i=1}^mk_i(a_i)$ on $\mathbb{P}^1$, for each $i\in\{1,...,m\}$, let ${\mathcal{C}}_i$ be the quasi-Hamiltonian $G$-space in Proposition \ref{k=k_i} with $k=k_i$. Then the symplectic space $\widetilde{\mathcal{M}}({\bf a})$ of monodromy/Stokes data for compatibly meromorphic connections $(V,\nabla,{\bf g})$ with irregular type ${\bf a}$ can be described as follows.
\begin{pro}[\cite{Boalch3} Lemma 3.1]
The symplectic space $\widetilde{\mathcal{M}}({\bf a})$ is isomorphic to the quasi-Hamiltonian quotient $\widetilde{\mathcal{C}}_1\circledast\cdot\cdot\cdot \circledast\widetilde{\mathcal{C}}_m\spr  G$, where $\circledast$ denotes the fusion product of two quasi-Hamiltonian $G$-manifolds.
\end{pro}

\subsection{Irregular Riemann-Hilbert correspondence}\label{sec53}
Let ${\bf a}$ be the data of a divisor $D=\sum k_i(a_i)$ and a fixed irregular type \ $\mathclap{^i}A^0$ at each $a_i$. The irregular Riemann-Hilbert correspondence, which depends on a choice of tentacles $\tau$ (see Definition 3.9 in \cite{Boalch2}), is a map $\nu$ from the global symplectic moduli space of meromorphic connections $\widetilde{\mathcal{M}^*}({\bf a})\cong (\widetilde{O}_1\times \cdot\cdot\cdot \widetilde{O}_m)\spr G$ to the symplectic space of monodromy data $\widetilde{\mathcal{M}}({\bf a})\cong (\widetilde{\mathcal{C}}_1\times\cdot\cdot\cdot\times\widetilde{\mathcal{C}}_m)\spr G$. In brief, the map arises as follows.

Let $(P,\nabla,{\bf g})$ be a compatibly framed meromorphic connection on a holomorphic trivial $G$-bundle $P$ with the irregular type ${\bf a}$. The chosen irregular type \ $\mathclap{^i}A^0$ canonically determines some directions at $a_i$ (‘anti-Stokes directions’), and we can consider the Stokes sectors at each $a_i$ bounded by two adjacent directions (and having some small fixed radius). Then the key fact is that, similar to the discussion in section \ref{defineC}, the framings ${\bf g}$ (and a choice of branch of logarithm at each pole) determine, in a canonical way, a choice
of solutions of the equation $\nabla F=0$ on each Stokes sector at each pole. Then along any path in the punctured sphere $\mathbb{P}^1\setminus \{a_1,... , a_m\}$ between two Stokes sectors, we can extend the two corresponding canonical solutions and obtain an element in $G$ by taking their ratio. The monodromy data of $(P,\nabla,{\bf g})$ is simply the set of all such elements in $G$, plus the exponents of formal monodromy, and thus corresponds to a point in the space of monodromy data $\widetilde{\mathcal{C}}_1\times\cdot\cdot\cdot\times\widetilde{\mathcal{C}}_m$. On the other hand, the triple $(P,\nabla,{\bf g})$ represents a point in $\widetilde{O}_1\times\cdot\cdot\cdot\widetilde{O}_m$. Therefore, it produces a map from $\widetilde{O}_1\times\cdot\cdot\cdot\widetilde{O}_m$ to $\widetilde{\mathcal{C}}_1\times\cdot\cdot\cdot\times\widetilde{\mathcal{C}}_m$ by taking the monodromy data of meromorphic connections $(P,\nabla, {\bf g})$. Furthermore, this map is $G$-equivariant and descends to give the irregular Riemann-Hilbert map $\nu$.
The main result of \cite{Boalch2} leads to:
\begin{thm}\label{RiemannHilbert}\rm{(\cite{Boalch2})}
The irregular Riemann-Hilbert map
\begin{eqnarray}\label{monodromymap}
\nu:(\widetilde{O}_1\times \cdot\cdot\cdot \widetilde{O}_m)\spr G \hookrightarrow
(\widetilde{\mathcal{C}}_1\circledast\cdot\cdot\cdot\circledast\widetilde{\mathcal{C}}_m)\spr G
\end{eqnarray}
associating monodromy/Stokes data to a meromorphic connection on a trivial $G$-bundle $P$ over $\mathbb{P}^1$ is a symplectic map (provided the symplectic structure on the right-hand side is divided by $2\pi i$).
\end{thm}

We will analyze the case where the meromorphic connections have one pole of order one and one pole of order two, and show that the irregular Riemann-Hilbert map $\nu$ gives rise to a local symplectic isomorphism from $(\Sigma,\omega)$ to $(\Sigma',\omega')$.
First we need the following two properties of the symplectic spaces $\widetilde{O}$ and quasi-Hamiltonian spaces $\widetilde{\mathcal{C}}$ with specific $k$'s.
\begin{pro}\label{isoO}
Let $\widetilde{O}_1$ and $\widetilde{O}_2$ be two copies of $\widetilde{O}$ with $k=1$ and $k=2$ respectively. Then the Hamiltonian quotient $\widetilde{O}_1\times\widetilde{O}_2\spr G$ is symplectic isomorphic to $(\Sigma,\frac{1}{2\pi i}\omega)$.
\end{pro}
\pf
By definition, $\widetilde{\mathcal{O}}_1=\{(g_1,x_1)\in G\times\frak g^*~|~g_1x_1g_1^{-1}\in \frak t''\}$ and $\widetilde{\mathcal{O}}_2=\{(g_2,A,x_2)\in G\times\frak g^*\times \frak g^*~|~{\rm Ad}_{g_2}A=A_0\}$, where $A_0\in \frak t_{\rm reg}$. Because $A$ is determined by $g_2$, $\widetilde{\mathcal{O}_2}$ is naturally isomorphic to $G\times\frak g^*$ by sending $(g_2,A,x_2)$ to $(g_2,x_2)$. Note that the moment map is
\begin{eqnarray*}
\mu:\widetilde{O}_1\times \widetilde{O}_2\longrightarrow\g^*; \ (g_1,x_1,g_2,x_2)\mapsto x_1+x_2.
\end{eqnarray*}
The submanifold $\mu^{-1}(0)$ is defined by $\mu^{-1}(0):=\{(g_1, x_1, g_2, -x_1)\in (G\times\frak g^*)^2~|~{\rm Ad}_{g_1}x_1\in \frak t''\}$. We have a subjective map
\begin{eqnarray*}
\iota:\mu^{-1}(0)\longrightarrow \Sigma; \ (g_1,x_1,g_2,-x_1)\mapsto (g_2g^{-1}_1,-2\pi i{\rm Ad}^*_{g_1}x_1)
\end{eqnarray*}
whose fibres are the $G$ orbits. Thus it induces a diffeomorphism from $\widetilde{O}_1\times\widetilde{O}_2\spr G$ to $\Sigma$.

Let us take two tangents $v_1,v_2$ to $\mu^{-1}(0)$ which at each point $(g_1,x_1,g_2,-x_1)$ take the forms $v_i=(0,{\rm Ad}_{g_1^{-1}}R_i,{\rm Ad}_{g_2^{-1}}X_i,-{\rm Ad}_{g_1^{-1}}R_i)$ for some $X_i\in\frak g, R_i\in\frak t^*$ and $i=1,2$ ($\frak g\cong T_{g_2}G$ via left multiplication).

Let $\omega_{\mu^{-1}(0)}$ be the restriction of the symplectic structure $\omega_{\widetilde{O}_1\times\widetilde{O}_2}$ on $\mu^{-1}(0)$.
Following the formula \eqref{omegaO}, we have that at $(g_1,x_1,g_2,-x_1)$,
\begin{eqnarray*}
\omega_{\mu^{-1}(0)}(v_1,v_2)&=&\omega_{\widetilde{O}_1}((0,{\rm Ad}_{g_1^{-1}}R_1),(0,{\rm Ad}_{g_1^{-1}}R_2))\nonumber\\
&&+\omega_{\widetilde{O}_2}(({\rm Ad}_{g_2^{-1}}X_1,-{\rm Ad}_{g_1^{-1}}R_1),({\rm Ad}_{g_2^{-1}}X_2,-{\rm Ad}_{g_1^{-1}}R_2))\nonumber\\
&=&\langle R_2,{\rm Ad}_{g_1g_2^{-1}}X_1\rangle-\langle R_1,{\rm Ad}_{g_1g_2^{-1}}X_2\rangle-\langle x_1,{\rm Ad}_{g_2^{-1}}([X_1,X_2])\rangle.
\end{eqnarray*}
On the other hand, a direct computation gives $\iota_*(v_i)=({\rm Ad}_{g_1g_2^{-1}}X_i,-2\pi i R_i)$ at $(g_2g_1^{-1},-2\pi i{\rm Ad}_{g_1}^*x_1)$, here $\frak g\cong T_{g_2g_1^{-1}}G$ via left multiplication. Formula \eqref{omegasigma} makes it transparent that at $(g_2g_1^{-1},-2\pi i {\rm Ad}^*_{g_1}x_1)\in \Sigma$,
\begin{eqnarray*}
\omega(\iota_*(v_1),\iota_*(v_2))&=&\omega({\rm Ad}_{g_2g_1^{-1}}X_1,-2\pi iR_1),({\rm Ad}_{g_2g_1^{-1}}X_2,-2\pi iR_2))\nonumber\\
&=&2\pi i\left(\langle R_2,{\rm Ad}_{g_1g^{-1}_2}X_1\rangle-\langle R_1,{\rm Ad}_{g_1g^{-1}_2}X_2\rangle-\langle x_1,Ad_{g^{-1}_2}([X_1,X_2])\rangle\right).
\end{eqnarray*}
Therefore, we have that $\frac{1}{2\pi i}\iota^*\omega=\omega_{\mu^{-1}(0)}$, i.e., $\iota$ induces a symplectic isomorphism between $\widetilde{O}_1\times\widetilde{O}_2\spr G$ and $(\Sigma,\frac{1}{2\pi i}\omega)$.
\qed
\\
\\
As for the Poisson Lie counterpart, we have
\begin{pro}\label{isoC}
Let $\widetilde{\mathcal{C}}_1$ and $\widetilde{\mathcal{C}}_2$ be two copies of $\widetilde{\mathcal{C}}$ with $k=1$ and $k=2$ respectively. Then the quasi-Hamiltonian quotient $\widetilde{\mathcal{C}}_1\circledast\widetilde{\mathcal{C}}_2\spr G$ is isomorphic to the symplectic submanifold $(\Sigma',\omega')$ of the double $\Gamma$.
\end{pro}
\pf
We assume that the Borels chosen at the first pole are opposite to those chosen at the second. Thus we have,
\begin{eqnarray*}
\widetilde{\mathcal{C}}_1=\{(h,(e^{\pi i\lambda^\vee},e^{-\pi i\lambda^\vee},\lambda^\vee))~|~h\in G,\lambda\in {\frak t}'\}, \ \ \ \ \ \widetilde{\mathcal{C}}_2=\{(C,(b_-,b_+,\Lambda))~|~\delta(b_{\pm})=e^{\pm\pi i\Lambda}\}.
\end{eqnarray*}
The moment map on $\widetilde{\mathcal{C}}_1\circledast \widetilde{\mathcal{C}}_2$ is $\mu=h^{-1}e^{-2\pi i\lambda^\vee}hC^{-1}b^{-1}_-b_+C$. Therefore the condition
$\mu=1$ becomes $Ce^{2\pi i{\rm Ad}_{h^{-1}}\lambda^\vee}C^{-1}=b_-^{-1}b_+$, where $B:={\rm Ad}_{h^{-1}}(\lambda^\vee)$. Recall that $\Sigma'$ is a submanifold of Lu-Weinstein symplectic double $\Gamma$,
\begin{eqnarray*}
\Sigma':=\{(g_1,(e^{\pi i\lambda^\vee},e^{-\pi i\lambda^\vee},\lambda^\vee),g_2,(b_-,b_+,\Lambda))\in\Gamma~|~\delta(b_{\pm})=e^{\pm\pi i\Lambda}, \ g_1e^{\pm \pi i \lambda^\vee}=b_{\pm}g_2\}.
\end{eqnarray*}
We have a surjective map from $\mu^{-1}(1)=\{(h,(e^{\pi i\lambda^\vee},e^{-\pi i\lambda^\vee},\lambda^\vee),C,(b_+,b_-,\Lambda))~|~e^{2\pi i {\rm Ad}_{h^{-1}}\lambda^{\vee}}=Cb_-^{-1}b_+C^{-1}\}$ to $\Sigma'$,
\begin{eqnarray*}
(h,(e^{\pi i\lambda^\vee},e^{-\pi i\lambda^\vee},\lambda^\vee),C,(b_+,b_-,\Lambda))\mapsto (Ch^{-1},(e^{\pi i\lambda^\vee},e^{-\pi i\lambda^\vee},\lambda^\vee),b_+^{-1}Ch^{-1}e^{\pi i\lambda},(b_-,b_+,\Lambda))
\end{eqnarray*}
whose fibres are precisely the $G$ orbits. Therefore, it induces an isomorphism from $\widetilde{\mathcal{C}}_1\circledast\widetilde{\mathcal{C}}_2\spr G$ to $\Sigma'$. An explicit formula for
the symplectic structure on $\Sigma'$ can be computed by using Theorem 3 of \cite{Anton1}. On the other hand we have
an explicit formula for the symplectic structure on $\widetilde{\mathcal{C}}_1\circledast\widetilde{\mathcal{C}}_2\spr G$. A straightforward calculation shows that these explicit formulas on each side agree.
\qed

\vspace{5mm}

It follows that the irregular Riemann-Hilbert map $\nu:(\widetilde{O}_1\times\widetilde{O}_2)\spr G\rightarrow (\widetilde{\mathcal{C}}_1\times \widetilde{\mathcal{C}}_2)\spr G$ induces a local symplectic isomorphism
\begin{eqnarray}\label{RHviaC}
\nu:\Sigma\rightarrow \Sigma'.
\end{eqnarray}
In the following, we will explicitly express $\nu$ by the connection map $C\in {\rm Map}(\frak g^*,G)$ defined in section \ref{defineC}. It enables us to finally prove Theorem \eqref{main}, as well as explain the relation between gauge equation \eqref{gaugeequation} and the symplectic nature of $\nu$.
\subsection{Irregular Riemann-Hilbert maps and gauge transformations of dynamical r-matrices}\label{GaugeviaRH}
According to \cite{Boalch2}, the Riemann-Hilbert map depends on some discrete data on the punctured Riemann sphere. To specify the map $\nu$ in \eqref{RHviaC}, we have to make a choice of tentacles.

Let us introduce coordinate $z$ to identify $\mathbb{P}^1$ with $\mathbb{C}\cup \infty$ and assume the divisor $D=1(a_1)+2(a_2)$ where $a_2=0$ and $a_1=\infty$. Then we consider the meromorphic connections $\nabla$ on a trivial holomorphic $G$-bundle $P$ on $\mathbb{P}^1$ with compatible framings ${\bf g}$, such that $(P,\nabla,{\bf g})$ have an irregular type $\frac{A_0}{z^2}$ at $0$, where $A_0\in\frak t_{\rm reg}$. Let us take a prior Stokes sector ${\rm Sect}_0$ between two adjacent Stokes rays at $0$, and make a choice of tentacles as follows.

\begin{itemize}

\item[\rm(i)] A choice of a point $p_2$ in ${\rm Sect}_0$ at $0$ and a point $p_1$ in ${\rm Sect}_0$ near $\infty$.

\item[\rm(ii)] A lift $\hat{p}_i$ of each $p_i$ to the universal cover of a punctured disc $D_i\backslash \{a_i\}$ containing $p_i$ for $i=1,2$.

\item[\rm(iii)] A base point $p_0$ which coincides with $p_1$.

\item[\rm(iv)] A contractible path $\gamma:[0,1]\rightarrow \mathbb{P}^1\setminus \{0,\infty\}$ in the punctured sphere, from $p_0$ to $p_1$.

\end{itemize}

Note that the chosen point $\hat{p}_2$ determines a branch of ${\rm log}z$ on ${\rm Sect}_0$. According to Section \ref{defineC}, let $C\in {\rm Map}(\g^*,G)$ be the connection map associated to $A_0\in \frak t_{\rm reg}$, the choice of ${\rm Sect}_0$ and the branch of ${\rm log}z$. Then we have
\begin{pro}\label{RHmap}
For the above choice of tentacles, the corresponding irregular Riemann-Hilbert map $\nu:(\widetilde{O}_1\times\widetilde{O}_2)\spr G\cong\Sigma\rightarrow (\widetilde{\mathcal{C}}_1\times \widetilde{\mathcal{C}}_2)\spr G\cong \Sigma'$ is given by
\begin{eqnarray*}
\nu(h,\lambda)=(C({\rm Ad}^*_h\lambda)h,e^{\lambda},u,u^*), \ \forall (h,\lambda)\in \Sigma,
\end{eqnarray*}
for certain $u\in G, u^*\in G^*$ satisfying $C({\rm Ad}^*_h\lambda)he^{\lambda}=u^*u$.
\end{pro}
\pf
Let $(P,\nabla,{\bf g}=(g_1,g_2))$ be a compatibly framed meromorphic connection with irregular type $\frac{A_0}{z^2}$ at $a_2$, where $g_1,g_2\in G$ and $A_0\in \frak t_{\rm reg}$.

Upon trivializing $V$, we assume $(P,\nabla,{\bf g})$ represents a point $(g_1,-x,g_2,A,x)\in \widetilde{O}_{1}\times\widetilde{O}_2$, which means that in the trivialization, $\nabla=d-(\frac{A}{z^2}+\frac{x}{z})dz$. Furthermore, the given irregular type $\frac{A_0}{z^2}$ of $\nabla$ in the compatible frame $g_2$ indicates that ${\rm Ad}_{g_2}A=A_0$. Using the convention in Section \ref{connectiondata}, let $F_0$ (resp. $F_\infty$) be the canonical solution of $\nabla_{A_0} F:=dF-(\frac{A_0}{z^2}+\frac{{\rm Ad}_{g_2}x}{z})Fdz=0$ at $0$ (resp. $\infty$). Due to the chosen frame ${\bf g}$, $\Phi_0=g_2^{-1}F_\infty g_2$, $\Phi_1=g_2^{-1}F_\infty g_2g_1^{-1}$ and $\Phi_2=g_2^{-1}F_0$ are the canonical solutions of $\nabla \Phi=0$ on a neighbourhood of $p_0=p_1$ and $p_2$ with respect to the compatible framing $1$, $g_1$ and $g_2$ respectively.
Then the monodromy data of $(P,\nabla,{\bf g})$
\begin{eqnarray*}
(C_1,(e^{\pi i\lambda^\vee},e^{-\pi i\lambda^\vee},\lambda^\vee),C_2,(b_-,b_+,\Lambda))\in\widetilde{\mathcal{C}}_1\times\widetilde{\mathcal{C}}_2\cong G\times e^{\frak t'}\times G\times G^*,
\end{eqnarray*}
is the set of connection matrices $C_i$ (the ratio of the canonical solutions $\Phi_i$ at $p_i$ with $\Phi_0$ at $p_0$ for $i=1,2$), as well as the Stokes data $(b_-,b_+)$ at $0$ and the formal monodromy at $0$, $\infty$. They are explicitly described as follows.
\\
\\
$\bullet$ along the path $\gamma$ in the punctured sphere $\mathbb{P}^1\setminus \{0, \infty\}$, we extend the two solutions $\Phi_0$ and $\Phi_2$, then $\Phi_2 C_2=\Phi_0$. Therefore we have $C_2=F_0^{-1}F_\infty g_2$. By definition, $F_0^{-1}F_\infty= C(2\pi i{\rm Ad}_{g_2}x)$ the connection matrix of $\nabla_{A_0}:=d-(\frac{A_0}{z^2}+\frac{{\rm Ad}_{g_2}x}{z})dz$;
\\
\\
$\bullet$ $p_0$, $p_1$ can be seen as connected by an identity path, thus $\Phi_1C_1=\Phi_0$. Therefore $C_1$ is equal to $g_1$, the ratio of the frame chosen at $p_0$ and $p_1$;
\\
\\
$\bullet$ $b_-,b_+$ at $0$ are the Stokes matrices of $\nabla_{A_0}$, which are determined by the monodromy relation \eqref{le:monodromy reln}.
\\

Therefore, the chosen tentacle determines a map $\nu':\mu^{-1}(0)\subset \widetilde{O}_1\times\widetilde{O}_2\rightarrow \mu'^{-1}(0)\subset\widetilde{\mathcal{C}}_1\times \widetilde{\mathcal{C}}_2$, which is given by
\begin{eqnarray*}
\nu'(g_1,-x,g_2,x)=(g_1, (e^{\pi i{\rm Ad}_{g_1}x^\vee},e^{-\pi i{\rm Ad}_{g_1}x^\vee},{\rm Ad}_{g_1}x^\vee),C(2\pi i{\rm Ad}_{g_2}x)g_2,(b_-,b_+,\Lambda)).
\end{eqnarray*}
This map $\nu'$ is $G$-equivariant and descends to the irregular Riemann-Hilbert map $\nu:\Sigma\rightarrow\Sigma'$
\begin{eqnarray*}
\nu(h,\lambda)=(C({\rm Ad}^*_h\lambda)h,e^{\lambda},u,u^*), \ \forall (h,\lambda)\in \Sigma.
\end{eqnarray*}
Here $u\in G, u^*\in G^*$ satisfy $C({\rm Ad}^*_h\lambda)he^{\lambda}=u^*u$, and we use the isomorphisms $\mu^{-1}(0)/G\cong\Sigma$ and $\mu'^{-1}(1)/G\cong \Sigma'$ constructed in Proposition \ref{isoO} and \ref{isoC} respectively.
\qed
\\
\\
Now we can finally give a proof of our main theorem.
\\

{\bf The proof of Theorem \ref{Cequation}.} By Theorem \ref{RiemannHilbert} and Proposition \ref{isoO}, the irregular Riemann-Hilbert map $\nu:(\Sigma,\omega)\rightarrow(\Sigma',\omega')$ is a symplectic map. By Proposition \ref{RHmap}, the map $\nu$ coincides with the local diffeomorphism $F_{C}:\Sigma\rightarrow \Sigma'$ defined in Section 2. Therefore, $F_C$ is a symplectic map. Eventually, by the equivalence given in Theorem \ref{equivalent}, the connection map $C$ is a solution of the equation \eqref{gaugeequation}. It finishes the proof.

\section{Vertex-IRF transformations, Drinfeld twists and connection maps}\label{twist}
In Section \ref{sec41} and \ref{sec42}, we recall respectively the following two results of Enriquez-Etingof-Marshall \cite{EEM}: the gauge equation \eqref{gaugeequation} is the semiclassical limit of a vertex-IRF transformation equation \cite{EN}; an admissible Drinfeld twist \cite{EH} killing an admissible associator gives rise to such a vertex-IRF transformation, and thus in turn gives a solution of equation \eqref{gaugeequation}. In Section \ref{sec43}, we study the gauge actions on the space of solutions of \eqref{gaugeequation} and on the space of admissible twists. This study enables us to prove Theorem \ref{gaugeasscl}, which states that in the case when $\g$ is semisimple, any solution of \eqref{gaugeequation} is a semiclassical limit of an admissible twist. Based on this result, in Section \ref{sec44}, we discuss the relation between Drinfeld twists and connection maps $C$ defined in Section \ref{connectiondata}.

\subsection{Vertex-IRF transformations and semiclassical limit}\label{sec41}
Let $(U(\g),m,\Delta,\varepsilon)$ denote the universal enveloping algebra of $\g$ with the product $m$, the coproduct $\Delta$ and the counit $\varepsilon$. Let $U(\g)\llbracket\hbar\rrbracket$ be the corresponding topologically free $\mathbb{C}\llbracket\hbar\rrbracket$-algebra. Let $\Phi=1+\frac{[t^{12},t^{23}]}{24}\hbar^2+O(\hbar^3)\in (U(\g)^{\hat{\otimes} 3})^\g\llbracket\hbar\rrbracket$ be a Drinfeld associator \cite{Drinfeld}, i.e., $\Phi$ satisfies the pentagon equation and the counit
axiom. 

Following the dynamical convention (see e.g. \cite{EE}), for a function $F:\g^*\rightarrow U(\g)^{\hat{\otimes} 2}$, we denote
\begin{eqnarray*}
F^{1,2}(x+{\hbar}h^{(3)}):=\sum_{N\ge 0}\frac{{\hbar}^N}{N!}\sum_{i_1,...,i_N}^n(\partial_{{\xi}^{i_1}}\cdot\cdot\cdot
\partial_{\xi^{i_N}}{F})(x)\otimes(e_{i_1}\cdot\cdot\cdot e_{i_N}),
\end{eqnarray*}
where $n={\rm dim}(\g)$, and $\{e_i\}_{i=1,...,n}$, $\{\xi^i\}_{i=1,...,n}$ are dual bases of $\g$ and $\g^*$. 

\begin{defi}\rm\cite{EE}\label{dynamicaltwist}
A function $J_d: \g^*\rightarrow U(\g)^{\hat{\otimes} 2}\llbracket\hbar\rrbracket$ is called a dynamical twist associated to $\Phi$ if
$J_d(x)=1+O(\hbar)$ is $\mathfrak{g}$-invariant and
\begin{eqnarray}\label{twist equation}
J_d^{12,3}J_d^{1,2}(x+\hbar h^{(3)})=\Phi^{-1}J_d^{1,23}(x)J_d^{2,3}(x).
\end{eqnarray}
Here $J_d^{2,3}(x)=1\otimes J_d(x)$, $J_d^{1,23}(x)=({\rm id}\otimes \Delta)(J_d(x))$, etc.
\end{defi}
Assume $(\Phi,J_d(x))$ satisfies the conditions in Definition \ref{dynamicaltwist}. Let $j(x):=(\frac{J_d(x)-1}{\hbar})$ mod $\hbar$, and $r(x):=j(x)-j(x)^{2,1}$. Then following \cite{EE}, $r(x)+\frac{t}{2}$ is a classical dynamical $r$-matrix, and ${J_d}(x)$ is called a dynamical twist quantization of $r(x)$.
Similarly, a constant twist $J_c\in U(\g)^{\hat{\otimes} 2}\llbracket\hbar\rrbracket$ is such that $J_c=1+\hbar{\frac{r}{2}}+O(\hbar^2)$ and
\begin{eqnarray}\label{constanttwist}
J_c^{12,3}J_c^{1,2}=\Phi^{-1}J_c^{1,23}J_c^{2,3}.
\end{eqnarray}
We say $J_c$ is a twist quantization of the classical r-matrix $r_0:=\frac{1}{2}(r-r^{2,1})$.

\begin{defi}\label{EN}{\rm \cite{EN}}
Let ${J_d}(x): \g^*\longrightarrow U(\g)^{\hat{\otimes} 2}\llbracket\hbar\rrbracket$ be a function with invertible values and $\rho: \g^*\longrightarrow U(\g)\llbracket\hbar\rrbracket$ a function with invertible values such that $\varepsilon(\rho(\lambda))=1$ ($\varepsilon$ is the counit). Set
\begin{eqnarray*}
J_d^{\rho}(x)=\Delta(\rho(x)){J_d}(x)\rho^1(x-\hbar h^{(2)})^{-1} \rho^2(x)^{-1},
\end{eqnarray*}
and call $\rho$ a vertex-IRF transformation from ${J_d}(x)$ to $J_d^\rho(x)$, where for $\rho^{-1}(x-\hbar h^{(2)})$ we use the dynamical convention.
\end{defi}

In particular, let $J_c$ (resp. $J_d(x)$) be a (resp. dynamical) twist quantization of $r_0$ (resp. $r_{\scriptscriptstyle \rm AM}$). Let $\rho(x)\in (U(\g)\hat{\otimes} \hat{S}^.\g)\llbracket\hbar\rrbracket$ be a formal vertex-IRF transformation which maps the dynamical twist ${J_d}(x)$ to the constant twist $J_c$.
This is to say
\begin{eqnarray}\label{IRFtransformation}
J_c=\Delta(\rho(x))J_d(x)\rho^1(x-h^{(2)})^{-1} \rho^2(x)^{-1}.
\end{eqnarray}
Then by comparing the coefficients of equation \eqref{IRFtransformation} up to the first order of $\hbar$, we have
\begin{pro}{\rm \cite{EEM}}\label{reduction}
The reduction modulo $\hbar$ of $\rho(x)$, denoted by $g(x)=\rho(x)|_{\hbar=0}$, belongs to ${\rm exp}(\g\otimes \hat{S}(\g))_{>0}$ (thus a formal map from $\g^*$ to $G$) and satisfies the equation $r_0^{g(x)}=r_{\scriptscriptstyle \rm AM}.$
\end{pro}

\subsection{Admissible Drinfeld twists and semiclassical limit}\label{sec42}
In \cite{EE}, the dynamical twist quantization $J_d(x)$ of $r_{\scriptscriptstyle \rm AM}$ is constructed by renormalizing (the inverses of) admissible Drinfeld associators. Later on in \cite{EEM}, the vertex-IRF transformation in \eqref{IRFtransformation} is obtained by renormalizing admissible Drinfeld twists. Let us recall the constructions.

{\bf Dynamical twist quantization of $r_{\scriptscriptstyle \rm AM}$ via admissible associators.}
Set $U':=U(\hbar \g\llbracket\hbar\rrbracket)\subset U(\g)\llbracket\hbar\rrbracket$. Note that $U'/\hbar U'=\hat{S}(\g)$, i.e., as a $\mathbb{C}\llbracket\hbar\rrbracket$-algebra, $U'$ is a deformation of $\hat{S}(\g)=\mathbb{C}\llbracket\g^*\rrbracket$. An associator $\Phi\in U(\g)^{\widehat{\otimes}3}\llbracket\hbar\rrbracket$ is called admissible (see \cite{EH}) if
$$\Phi\in 1+\frac{\hbar^2}{24}[t^{1,2},t^{2,3}]+O(\hbar^3),\ \ \ \ \ \hbar {\rm log}(\Phi)\in (U')^{\widehat{\otimes}3}.$$
Given an admissible associator $\Phi\in U(\g)^{\hat{\otimes} 3}\llbracket\hbar\rrbracket$, we identify the third component $U(\g)$ of this tensor cube with $S^.(\g)$ via the symmetrization (PBW) isomorphism. This identification enables us to view $\Phi$ as a formal function $\Phi^{-1}(x): \g^*\rightarrow U(\g)^{\hat{\otimes} 2}\llbracket\hbar\rrbracket$. Furthermore, the $\hbar$-adic valuation properties of $\Phi$ gives us a well-defined element $\Phi^{-1}(\hbar^{-1}x)$ in $(U(\g)^{\hat{\otimes} 2}\hat{\otimes}\hat{S}(\g))\llbracket\hbar\rrbracket$. Following \cite{EH}, any universal Lie associator gives rise to an admissible associator.

\begin{thm}\rm{\cite{EE}}\label{EEthm}
Assume that $\Phi$ is the image in $U(\g)^{\hat{\otimes} 3}\llbracket\hbar\rrbracket$ of a universal Lie associator. Let $J_d(x):=\Phi^{-1}(\hbar^{-1}x)$, where $\Phi^{-1}(x)$ is regarded as an element of $(U(\g)^{\hat{\otimes} 2}\hat{\otimes} \hat{S}(\g))\llbracket\hbar\rrbracket$. Then

$(1)$ $J_d(x)$ is a formal dynamical twist. More precisely, $J_d(x)=1+\hbar j(x)+O(\hbar^2)\in (U(\g)^{\hat{\otimes} 2}\hat{\otimes} \hat{S}(\g))\llbracket\hbar\rrbracket$, is a series in nonnegative powers of $\hbar$ and satisfies the dynamical twist equation.

$(2)$ $J_d(x)$ is a twist quantization of the Alekseev-Meinrenken dynamical $r$-matrix, that is $r_{\scriptscriptstyle \rm AM}=j(x)-j(x)^{2,1}$.

$(3)$ If $\Phi_{\rm KZ}$ is the Knizhnik-Zamolodchikov associator, then $J_d(x)$ is holomorphic on an open set and extends meromorphically to the whole $\g^*$.
\end{thm}

{\bf Vertex-IRF transformations via admissible twists.}
For an admissible associator $\Phi$, there exists a twist killing $\Phi$ (see \cite{Drinfeld}\cite{EK}), and according to \cite{EH}, this twist can be made admissible by a suitable gauge transformation. The resulting twist $J\in U(\g)^{\hat{\otimes} 2}\llbracket\hbar\rrbracket$ satisfies $J=1-\hbar\frac{r}{2}+O(\hbar^2)$, $\hbar {\rm log}(J)\in U'^{\widehat{\otimes}2}, (\varepsilon\otimes {\rm id})(J)=({\rm id} \otimes \varepsilon)(J)=1$, and
\begin{equation}\label{Drinfeldtwist}
\Phi=(J^{2,3} J^{1,23})^{-1} J^{1,2} J^{12,3}.
\end{equation}
Let $J_d(x)=\Phi(\hbar^{-1}x)$ be the dynamical twist in Theorem \ref{EEthm}, then the admissible twist $J$ is used in \cite{EEM} to construct IRF-transformations satisfying \eqref{IRFtransformation} as follows.

First, let us identify the second component $U(\g)$ of $J$ with $\mathbb{C}(\g^*)$ via PBW isomorphism $S^.(\g)
\cong U(\g)$, and regard $J$ as a formal function from $\g^*$ to $U(\g)\llbracket\hbar\rrbracket$, denoted by $J(x)$. Let $\rho(x):=J(\hbar^{-1}x)\in {\rm Map}(\g^*,U(\g)\llbracket\hbar\rrbracket)$ denote the corresponding renormalization by sending $x\in\g^*$ to $\hbar^{-1} x$. Then if we identify the third component $U(\g)$ of the tensor cube with $\mathbb{C}(\g^*)$ in equation \eqref{IRFtransformation} and renormalize the resulting formal maps from $\g^*$ to $U(\g)^{\hat{\otimes} 2}$ by sending $x\in \g^*$ to $\hbar x$, the equation \eqref{Drinfeldtwist} becomes
\begin{eqnarray*}
J^{-1}=\Delta(\rho(x))J_d(x)\rho^1(x-h^{(2)})^{-1} \rho^2(x)^{-1}
\end{eqnarray*}
(Here $J_d(x):=\Phi^{-1}(\hbar^{-1}x)$ is the dynamical twist as in Theorem \ref{EEthm}). One checks that $J_c:=J^{-1}$ satisfies \eqref{constanttwist} (and thus is a constant twist). Therefore, the admissible Drinfeld twist $J$ gives rise to a vertex-IRF transformation between the dynamical twist $J_d(x)$ and the constant twist $J_c=J^{-1}$.

Now given an admissible Drinfeld twist $J\in U(\g)\llbracket\hbar\rrbracket\hat{\otimes}U'$, we denote its reduction mod $\hbar$ by $g(x)\in U(\g)\hat{\otimes}\hat{S}(\g)= U(\g)\llbracket \g^*\rrbracket$, which is a formal series on $\g^*$ with coefficients in $U(\g)$. By the discussion above and Proposition \ref{reduction}, $g(x)$ is actually in ${\rm Map}_0(\g^*,G)$, and satisfies equation \eqref{gaugeequation}. That is
\begin{pro}\rm\cite{EEM}
The semiclassical limit $g(x)$ of an admissible Drinfeld twist $J$ satisfies equation \eqref{gaugeequation}.
\end{pro}
For the case when $\g$ is semisimple, we will prove that the inverse is also true, i.e.,
given any solution
$g(x)$ of \eqref{gaugeequation}, there exists an admissible Drinfeld twist $J$ whose classical limit is $g(x)$. In order to prove this, we need to consider the gauge action on the set of Drinfeld twists and on the space of solutions of \eqref{gaugeequation}.

\subsection{Gauge actions}\label{sec43}
{\bf Gauge actions on Drinfeld twists.} Recall that $U'=U(\hbar \g\llbracket\hbar\rrbracket)$. Let $U'_0:={\rm Ker}(\varepsilon)\cap U'$. Then
$V:=\{u_\hbar\in\hbar^{-1}U'_0\subset U(\g)\llbracket\hbar\rrbracket\}~|~u_\hbar=O(\hbar)\}$ is a Lie subalgebra for
the commutator. One checks that $e^{u_\hbar}\ast J:=(e^{u_\hbar})^1(e^{u_\hbar})^2J(\Delta(e^{u_\hbar}))^{-1}$ is a solution of
\eqref{Drinfeldtwist} if $J$ is. Then the infinitesimal gauge action of $V$ on the set of Drinfeld twists is given by $$\delta_{u_\hbar}(J)={u_\hbar}^1J+{u_\hbar}^2J-J{u_\hbar}^{12}, \ \ {u_\hbar}\in V.$$ Note that $V/\hbar V=(\hat{S}(\g)_{>1},\{-,-\})$. Let $u\in \hat{S}(\g)_{>1}$ be the reduction modulo $\hbar$ of ${u_\hbar}\in V$, and $g\in {\rm exp}(\g\otimes \hat{S}(\g)_{>0})$ the semiclassical limit of $J$.
Then the reduction modulo $\hbar$ of the infinitesimal gauge action is
\begin{eqnarray}\label{classicalaction}
\delta_u(g)=\{1\otimes u, g\}-g\cdot du,\end{eqnarray}
where $du:=e_i\otimes \frac{\partial u}{\partial\xi^i}\in \g\otimes \hat{S}(\g)$ for an orthogonal basis $\{e_i\}$ of $\g$ and the corresponding coordinates $\{\xi^i\}$ on $\g^*$.
This infinitesimal gauge action has a geometric description as follows.
\\
\\
{\bf Gauge actions on the space of solutions of \eqref{gaugeequation}.} Recall that ${\rm Map}_0(\g^*,G)$ is the space of formal maps $g:\g^*\rightarrow G$ such that $g(0)=1$. Let us introduce a group structure on ${\rm Map}_0(\g^*,G)$, defined by $(g_1\ast g_2)(x):=g_2({\Ad}^*_{g_1(x)}x)g_1(x)$. Then there is a natural group homomorphism \[{\rm Map}_0(\g^*,G)\rightarrow {\rm Diff}(\g^*),\]
which maps $g\in {\rm Map}_0(\g^*,G)$ to the diffeomorphism $g\cdot x={\rm Ad}_{g(x)}x, \ \forall x\in\g^*$. Let us take the subgroup ${\rm Map}_0^{ham}(\g^*,G)$ whose elements, under the above group homomorphism, correspond to Poisson isomorphisms on $\g^*$ (equipped with its canonical linear Poisson structure). Explicitly, the elements $g$ of ${\rm Map}_0^{ham}(\g^*,G)$ are such that (we use the same convention as in \eqref{gaugeequation})
\[g_1^{-1}d_2(g_1)-g_2^{-1}d_1(g_2)+\langle {\rm id}\otimes {\rm id}\otimes x,[g_1^{-1}d_3(g_1),g_2^{-1}d_3(g_2)]\rangle=0.\]
Then it is direct to check that ${\rm Map}^{ham}_0(\g^*,G)$ is a prounipotent Lie group with Lie algebra $\{\alpha\in \g\otimes \hat{S}(\g)_{\ge 1}~|~{\rm Alt}(d\alpha)=0\}$. This Lie algebra is isomorphic to $(\hat{S}(\g)_{> 1},\{-,-\})$ under the map $d:u\rightarrow du\in\g\otimes \hat{S}(\g)_{\ge 1}$, for all $f\in\hat{S}(\g)_{> 1}$.

The right action of ${\rm Map}_0(\g^*,G)$ on itself restricts to an action of ${\rm Map}_0^{ham}(\g^*,G)$ on the space of solutions of equation \eqref{gaugeequation}. The action of $\alpha\in {\rm Map}_0^{ham}(\g^*,G)$ on a solution $g$ is given by $(\alpha\ast g)(x)=g({\Ad}^*_{\alpha(x)}x)\alpha(x)$.
The infinitesimal of this action is that each $u\in (\hat{S}(\g)_{> 1},\{-,-\})$ (the Lie algebra of ${\rm Map}_0(\g^*,G)$) acts as vector fields on the space of solutions by
\begin{eqnarray}\label{classicalaction1}
g^{-1}\delta_u(g)=\langle {\rm id}\otimes {\rm id}\otimes x, [d_3(u_2),g^{-1}_{12}d_3g_{12}]\rangle -du\in \g\otimes \hat{S}(\g)_{\ge 0}.
\end{eqnarray}
It coincides with the infinitesimal gauge action \eqref{classicalaction}. Therefore we have a commutative diagram of gauge actions and taking semiclassical limit
$$
\begin{CD}
J @> e^{u_\hbar}>> e^{u_\hbar}\ast J \\
@V s.c.l VV @V s.c.l VV \\
g @> e^u >> e^u\ast g
\end{CD}
$$
Here we assume the semiclassical limit of $J$ (resp. $u_\hbar\in V$) is $g$ (resp. $u\in \hat{S}(\g)_{> 1}$), and $e^u$ is seen as an element in ${\rm Map}_0^{ham}(\g^*,G)$ under the Lie algebra isomorphism $\hat{S}(\g)_{> 1}\cong {\rm Lie}({\rm Map}_0^{ham}(\g^*,G))$. This fact enables us to prove the following theorem.

\begin{thm}\label{gaugeasscl}
For a semisimple Lie algebra $\g$, given any formal solution $g\in {\rm Map}_0(\g^*,G)$ of equation \eqref{gaugeequation}, $r_0^g=r_{\scriptscriptstyle \rm AM}$, there exists an admissible Drinfeld twist $J$ whose semiclassical limit is $g$.
\end{thm}
\pf
Let $J'$ be an admissible Drinfeld twist and $g'(x)$ its semiclassical limit (thus a solution of
\eqref{gaugeequation}).
Following \cite{EEM}, ${\rm Map}_0^{ham}(\g^*,G)$ acts simply and transitively on the space of solutions of \eqref{gaugeequation}.
Therefore, the two solutions $g$ and $g'$ are related by a map $\alpha\in {\rm Map}^{ham}_0(\g^*,G)$, i.e., $g(x)=g'\ast \alpha$. Let us assume $u\in \hat{S}(\g)_{> 1}$ (Lie algebra of ${\rm
Map}^{ham}_0(\g^*,G)$) is such that $e^u=\alpha$, and then take an element $u_\hbar \in V\subset U(\g)\llbracket\hbar\rrbracket$ whose reduction modulo $\hbar$ is $u$. The gauge action of $e^{u_\hbar}$ on $J'$ provides a new admissible twist $J:= e^{u_\hbar}\ast J'$. Furthermore the above commutative diagram verifies $g(x)=J(\hbar^{-1}x)|_{\hbar=0}$ (regard $J$ as a formal function from $\g^*$ to $U(\g)\llbracket\hbar\rrbracket$), i.e., the semiclassical limit of $J$ is $g(x)$.
\qed
\subsection{Drinfeld twists and connection maps}\label{sec44}
In Section \ref{connectiondata}, we have shown that any connection map $C\in {\rm Map}_0(\g^*,G)$ is a solution of \eqref{gaugeequation}. Note that Taylor expansion at the origin takes $C$ to a formal
map. Thus, as an immediate consequence of Theorem \ref{gaugeasscl}, we have the following statement:
\begin{cor} Assume $\Phi$ is the image in $U(\g)^{\hat{\otimes}3}$ of a universal Lie associator. Then for any connection map $C\in {\rm Map}_0(\g^*,G)$, there exists an admissible Drinfeld twist $J$ killing the associator $\Phi$ whose semiclassical limit is $C$.
\end{cor}

In particular, let $\Phi$ be the Knizhnik-Zamolodchikov (KZ) associator $\Phi_{KZ}$, which is the monodromy from $1$ to $\infty$ of the KZ equation on $\mathbb{P}^1$ with three simple poles at $0$, $1$, $\infty$. Naively, the confluence of two simple poles at $0$ and $1$ in the KZ equation leads to a degree two pole, while the monodromy representing KZ associator becomes the connection matrix $C_{\hbar}$ for an irregular Riemann-Hilbert problem. Then Theorem \ref{Cequation} and the above corollary indicate that the monodromy $C_{\hbar}$ may give a certain Drinfeld twist killing $\Phi_{KZ}$. Indeed, an explicit construction of the Drinfeld twist along this way is given in \cite{TL}. See \cite{TLXu} for a further discussion and the relation to the present paper.

\appendix

\section{Appendix: Proof of Theorem \ref{equivalent}}
In this section, we will study in details the symplectic submanifold $\Sigma'$ of Lu-Weinstein symplectic double groupoid $\Gamma$ and then give a proof of Theorem \ref{equivalent}. According to Section 2, given a quasitriangular Lie bialgebra $(\frak g,r)$, $(\Gamma,\pi_{\Gamma})$ is the set
\begin{eqnarray*}
\Gamma=\{(h,h^*,u,u^*)~|~h,u\in G,h^*,u^*\in G^*, hh^*=u^*u\}
\end{eqnarray*}
with the unique Poisson structure $\pi_{\Gamma}$ such that the local diffeomorphism $(\Gamma,\pi_{\Gamma})\rightarrow (D,\pi_D)$: $(h,h^*,u,u^*)\mapsto hh^*$ is a Poisson map ($D$ is the double Lie group). Then the submanifold $\Sigma'$ takes the form
\begin{eqnarray*}
\Sigma'=\{(h,h^*,u,u^*)\in \Gamma~|~h^*\in e^{\frak t'}\subset G^*\}.
\end{eqnarray*}
\begin{pro}
$\Sigma'$ is a symplectic submanifold of the Lu-Weinstein symplectic double $(\Gamma,\pi_{\Gamma})$.
\end{pro}
\pf An explicit formula for the restriction of symplectic $2$-form on $\Sigma'\in\Gamma$ can be computed by using Theorem 3 of \cite{AM}. One can check directly that it is symplectic. \qed
\\

Thus $\Sigma'$ inherits a symplectic structure $\omega'$. We denote by $\pi'$ the corresponding Poisson bivector. Note that the inclusion $(\Sigma',\pi')\hookrightarrow (\Gamma,\pi_{\Gamma})$ and the dressing transformation map $(\Gamma,\pi_{\Gamma})\rightarrow (G^*,\pi_{G^*})$; $(h,h^*,u,u^*)\mapsto d_h(h^*)$ are Poisson, and so is their composition. Thus we have
\begin{pro}\label{Poissonprojection}
The map
\begin{eqnarray*}
P':(\Sigma',\pi')\rightarrow (G^*,\pi_{G^*}); \ (h,e^{\lambda},u,u^*)\mapsto d_he^{\lambda}
\end{eqnarray*}
is a Poisson map.
\end{pro}

To simplify the notation, we take a local model of $(\Sigma',\pi')$ as follows. Recall that we have the local diffeomorphism
\begin{eqnarray*}
\Sigma'\rightarrow G\times e^{\frak t'}; \ (h,e^{\lambda},u,u^*)\mapsto (h,e^{\lambda}).
\end{eqnarray*}
We will take $G\times e^{\frak t'}$ as a local model of $(\Sigma',\pi)$ with the induced Poisson tensor, denoted also by $\pi'$. Generally, $\pi'$ is only defined on a dense subset of $G\times e^{\frak t'}$, however this is enough for our purpose.

Let $T$ act on $G\times e^{\frak t'}$ by $t\cdot (h,e^{\lambda})=(ht,e^{\lambda})$. The fibres of the map $P':G\times\frak t'\rightarrow G^*$, $(h,e^\lambda)\mapsto d_h(e^\lambda)$ are precisely the $T$-orbits. Thus a general $1$-form on $G\times e^{\frak t'}$ takes the form $P'^*(\beta)+\hat{\eta}$, where $\beta\in \Omega^1(G^*)$, $\eta\in \frak t^*\subset \frak g^*$ and at each point $(h,e^{\lambda})$, $\hat{\eta}:=(l_{h^{-1}}\circ r_{e^{-\lambda}})^*\eta$.
\begin{pro}\label{transition}
At each point $(h,e^{\lambda})$, $\pi'$ is given for any 1-forms $\phi_1:=P'^*(\beta_1)+\hat{\eta}_1$, $\phi_2:=P'^*(\beta_2)+\hat{\eta}_2$ by
\begin{eqnarray}\label{pi'}
\pi'(h,e^{\lambda})(\phi_1,\phi_2)&=&\pi_{G^*}(d_he^{\lambda})(\beta_1,\beta_2)+\langle X_1,\eta_2\rangle-\langle X_2,\eta_1\rangle\nonumber\\
&+&(l_{h^{-1}}\pi_G(h))(\eta_1,\xi_2)-(l_{h^{-1}}\pi_G(h))(\eta_2,\xi_1)+(l_{h^{-1}}\pi_G(h))(\eta_1,\eta_2),
\end{eqnarray}
where $\xi_i+X_i\in \frak g^*\otimes\frak g$ is the pull back of $P'^*(\beta_i)$ under $l_{h^{-1}}\circ r_{e^{-\lambda}}$ for $i=1,2$.
\end{pro}
\pf
Following \cite{Lu1}, if $m\in D$ ( the double Lie group) can be factored as $m=hu$ for some $h\in G$ and $u\in G^*$ (locally it is always the case), then explicit formula for $\pi_D$ is given by
\begin{eqnarray*}
&&((l_{h^{-1}}\circ r_{u^{-1}})(\pi_D))(m)(\xi_1+X_1,\xi_2+X_2)\nonumber\\
&=&\langle X_1,\xi_2\rangle-\langle X_2,\xi_1\rangle+(l_{h^{-1}}\pi_G(h))(\xi_1,\xi_2)+(r_{u^{-1}}\pi_{G^*}(u))(X_1,X_2)
\end{eqnarray*}
for $\xi_1+X_1,\xi_2+X_2\in \frak g^*\oplus\frak g$.

On one hand, Proposition \ref{Poissonprojection} gives that
\begin{eqnarray*}
\pi'(P'^*(\beta_1),P'^*(\beta_2))=\pi_{G^*}(\beta_1,\beta_2),
\end{eqnarray*}
for any $\beta_1,\beta_2\in\Omega^1(G^*)$.

On the other hand, let us consider the one form taking the form of $\hat{\eta}:=l^*_{h^{-1}}(r^*_{e^{-\lambda}}\eta)$, $\eta\in\frak t^*\subset \frak g^*$. From the expression of $\pi_D$, we see that $\pi^{\sharp}_D(he^{\lambda})(\hat{\eta})$ is tangent to $G\times e^{\frak t'}$ at $(h,e^{\lambda})$. Thus
\begin{eqnarray*}
\pi'(\hat{\eta}_1+P'^*(\beta_1),\hat{\eta}_2)&=&\pi_D(\hat{\eta}_1+d^*(\beta_1),\hat{\eta}_2)|_{G\times e^{\frak t'}}\nonumber\\
&=&\langle X_1,\eta_2\rangle+l_{h^{-1}}\pi_G(\xi_1,\eta_2)+l_{h^{-1}}\pi_G(\eta_1,\eta_2)
\end{eqnarray*}
where $\xi_1+X_1\in \frak g^*\otimes\frak g$ is the pull back of $P'^*(\beta_1)$ under $l_{h^{-1}}\circ r_{e^{-\lambda}}$. The above two identities indicate the expression \eqref{pi'} of $\pi'$.
\qed
\\

In the following, we will give a description of the Poisson space $(G\times e^{\frak t'},\pi')$ via dynamical $r$-matrices.
Let us define a bivector field on $G\times\frak t'$ which at each point $(h,\lambda)$ takes the form
\begin{eqnarray*}
\pi_r(h,\lambda)=l_h(t_i)\wedge \frac{\partial}{\partial t^i}+l_h(({\rm id}\otimes {\rm ad^{-1}_{\lambda^\vee}})(t))+l_h(r_{\scriptscriptstyle \rm AM}(\lambda))-r_h(r_0)
\end{eqnarray*}
where $t\in S^2(\frak g)^{\frak g}$ is the Casimir element, $\{t_i\}$ is a basis of $\frak t$ and $\{t^i\}$ the corresponding coordinates on $\frak t^*$ and at any point $x\in\frak g$, ${\rm ad}^{-1}_x:\frak g\rightarrow \frak g$ is the trivial extension of the map ${\rm ad}^{-1}_x:{\frak g}^\perp_x \rightarrow {\frak g}^\perp_p\subset \frak g$ corresponding to the decomposition $\frak g=\frak g_x\oplus{\frak g}^\perp_x$. Here $\frak g_x$ is the isotropic subalgebra of $\frak g$ at $x$ and $\frak g_x^\perp$ its complement with respect to the nondegenerate bilinear form. Using the fact that $r_{\scriptscriptstyle \rm AM}$ satisfies the dynamical Yang-Baxter equation, one checks that $\pi_r$ is a Poisson bivector.
\begin{pro}\label{local}
The image of $\pi_r$ under the diffeomorphism $\Psi:G\times\frak t'\rightarrow G\times e^{\frak t'}$, $(h,\lambda)\mapsto (h,e^{\lambda})$ coincides with $\pi'$.
\end{pro}

Before giving a proof, we introduce the following lemma. Let us consider the Semenov-Tian-Shansky (STS) Poisson tensor on $\frak g^*$ defined by
\begin{eqnarray*}
\pi_{\scriptscriptstyle \rm STS}(x)(df,dg)=\langle df(x)\otimes dg(x), {\rm ad}_x\otimes\frac{1}{2}{\rm ad}_x{\rm coth}(\frac{1}{2}{\rm ad}_x)(t)-\otimes^2{\rm ad}_x(r_0)\rangle,
\end{eqnarray*}
for any $f, g\in C^{\infty}(\g^*)$. We denote by $L,R$ the group morphisms corresponding to the Lie algebra morphisms $L,R:\frak g^*\rightarrow \frak g$
\begin{eqnarray*}
L(x):=(x\otimes {\rm id})(r), \ \ \ \ \ \ R(x):=-(x\otimes {\rm id})(r^{2,1}) \ \ \ \ \ \ \forall \ x\in \frak g^*.
\end{eqnarray*}
\begin{lem}\rm{\cite{FM}}\label{STSPoisson}
The map $I:(\frak g^*,\pi_{\scriptscriptstyle \rm STS})\rightarrow (G^*,\pi_{G^*})$ determined by $e^{x^{\vee}}=L(I(x))^{-1}R(I(x))$ for any $x\in\frak g^*$, is a Poisson map.
\end{lem}

{\bf Proof of Proposition \ref{local}.}
Recall that (from the discussion above Proposition \ref{transition}) a general $1$-form on $G\times e^{\frak t'}$ takes the form $P'^*(\beta)+\hat{\eta}$, where $\beta\in\Omega^1(G^*)$ and $\eta\in\frak t^*\subset \frak g^*$. Thus we only need to prove that $\Psi_*(\pi_r)(\hat{\eta},\cdot)=\pi'(\hat{\eta},\cdot)$ and $\Psi_*(\pi_r)(P'^*(\beta),\cdot)=\pi'(P'^*(\beta),\cdot)$. First note that at each point $(h,\lambda)$ (by the equivariance of $r_{\scriptscriptstyle \rm AM}$)
\begin{eqnarray*}
l_h(r_{\scriptscriptstyle \rm AM}(\lambda)+({\rm id}\otimes {{\rm ad}^{-1}_{\lambda^\vee}})(t))=r_h(r_{\scriptscriptstyle \rm AM}(x)+({\rm id}\otimes {{\rm ad}^{-1}_{x^\vee}})(t))=r_h(({\rm id}\otimes{\rm coth}(\frac{1}{2}{\rm ad}_{x^\vee})(t)),
\end{eqnarray*}
where $x={\rm Ad}^*_h\lambda\in \frak g^*$. By the definition of the map $P$, a direct calculation gives that
\begin{eqnarray*}
\pi_r(h,\lambda)(P^*(\alpha_1),P^*(\alpha_2))=({\rm ad}^*_{x^\vee}\otimes\frac{1}{2}{\rm ad}^*_{x^\vee}{\rm coth}(\frac{1}{2}{\rm ad}^*_{x^\vee})(t)-\otimes^2{\rm ad}^*_{x^\vee}(r_0)(\alpha_1,\alpha_2),
\end{eqnarray*}
where $(h,\lambda)\in G\times\frak t'$ and $x={\rm Ad}^*_h\lambda$. In other words, \begin{eqnarray*}
\pi_r(h,\lambda)(P^*(\alpha_1),P^*(\alpha_2))=\pi_{\scriptscriptstyle \rm STS}(x)(\alpha_1,\alpha_2).
\end{eqnarray*}
On the other hand, we have the following commutative diagram
$$
\begin{CD}
G\times\frak t' @> \Psi>> G\times e^{\frak t'} \\
@V P VV @V P' VV \\
\frak g^* @> I >> G^*
\end{CD},
$$
where $I:(\frak g^*,\pi_{\scriptscriptstyle \rm STS})\rightarrow (G^*,\pi_{G^*})$ is the local Poisson isomorphism in Lemma \ref{local}.
Thus $P^*(I^*(\beta_i))=\Psi^*(P'^*(\beta_i))$ for any $\beta_i\in\Omega^1(G^*)$, $i=1,2$. Therefore,
\begin{eqnarray*}
\Psi_*\pi_r(P'^*(\beta_1),P'^*(\beta_2))&=&\pi_r(P^*(I^*(\beta_1)),P^*(I^*{\beta_2}))=\pi_{\scriptscriptstyle \rm STS}(I^*(\beta_1),I^*(\beta_2)),\nonumber \\
\pi'(P'^*(\beta_1),P'^*(\beta_2))&=&\pi_{G^*}(\beta_1,\beta_2).
\end{eqnarray*}
Now Lemma \ref{STSPoisson} says that $\pi_{\scriptscriptstyle \rm STS}(I^*(\beta_1),I^*(\beta_2))=\pi_{G^*}(\beta_1,\beta_2)$, thus we have that \begin{eqnarray*}
\Psi_*\pi_r(P'^*(\beta_1),P'^*(\beta_2))=\pi'(P'^*(\beta_1),P'^*(\beta_2)).
\end{eqnarray*}

For the remaining part, by the definition of the diffeomorphism $\Psi$ and the expression of $\pi_G=l_h(r_0)-r_h(r_0)$, one can easily get that
\begin{eqnarray*}
\Psi_*(\pi_r)(P'^*(\beta_1)+\hat{\eta_1},\hat{\eta})=\langle X_1,\eta\rangle+l_{h^{-1}}\pi_G(\xi_1,\eta)+l_{h^{-1}}\pi_G(\eta_1,\eta),
\end{eqnarray*}
where $\xi_1+X_1\in \frak g^*\otimes\frak g$ is the pull back of $P'^*(\beta_1)$ under $l_{h^{-1}}\circ r_{e^{-\lambda}}$ and $\eta,\eta_1\in \frak t'$. By comparing with the expression of $\pi'$, we have that $\Psi_*(\pi_r)(\hat{\eta},\cdot)=\pi'(\hat{\eta},\cdot)$ for any $\eta\in \frak t^*$.

Eventually, we prove that $\Psi_*(\pi_r)(P'^*(\beta)+\hat{\eta},\cdot)=\pi'(P'^*(\beta)+\hat{\eta},\cdot)$ for any $\beta\in\Omega^1(G^*)$, $\eta\in \frak t^*\subset \frak g^*$. That is, the image of $\pi_r$ under the diffeomorphism $\Psi$ coincides with $\pi'$.
\qed
\\

In other words, we have a local symplectic isomorphism $\Psi:(G\times \frak t', \pi_r)\rightarrow (\Sigma',\pi')$, $(h,\lambda)\mapsto (h,e^{\lambda},u,u^*)$, where $u\in G, u^*\in G^*$ are determined by the identity $he^{\lambda}=u^*u$.

For the Poisson tensor $\pi$ corresponding to the symplectic form $\omega$ on $G\times \frak t'$, we have
\begin{pro}
The Poisson tensor $\pi$ takes the form
\begin{eqnarray*}
\pi(h,\lambda)=l_h(t_j)\wedge \frac{\partial}{\partial t^j}+l_h({\rm id}\otimes({\rm ad^{-1}_{\lambda^\vee}})(t))
\end{eqnarray*}
where $\{t_j\}$ is a basis of $\frak t$ and $\{t^j\}$ the corresponding coordinates on $\frak t^*$.
\end{pro}
After this preliminary work, we can give a proof of Theorem \ref{equivalent}. Following Proposition \ref{local}, $(G\times\frak t',\pi_r)$ is locally isomorphic to $(\Sigma',\pi')$. Therefore we can take $(G\times\frak t',\pi_r)$ as a local model of $(\Sigma',\pi')$ and then the map defined by \eqref{DiffFg} becomes $F_g:\Sigma=G\times\frak t'\rightarrow G\times \frak t'$, $(h,\lambda)\mapsto(g(x)h,\lambda)$, where $x={\rm Ad}^*_h\lambda$. Theorem \ref{equivalent} is thus equivalent to
\begin{thm}
$F_g:(G\times\frak t',\pi)\rightarrow (G\times\frak t',\pi_r)$ is a Poisson map if and only if $g\in {\rm Map}(\frak g^*,G)$ satisfies the gauge transformation equation \eqref{gaugeequation}, $r_0^{g}=r_{\scriptscriptstyle \rm AM}$.
\end{thm}
\pf
We only need to show that ${F_g}_*\pi=\pi_r$ is equivalent to the equation $r_0^g=r_{\scriptscriptstyle \rm AM}$. By comparing the expressions of $\pi$ and $\pi_r$, we have
\begin{eqnarray*}
\pi_r(h,\lambda)=\pi(h,\lambda)+l_h(r_{\scriptscriptstyle \rm AM}(\lambda)-\otimes^2{\rm Ad}_{h^{-1}}(r_0)).
\end{eqnarray*}
At any point $(h,\lambda)\in G\times\frak t'$, $F_g(h,\lambda)=(g(x)h,\lambda)$ where $x:={\rm Ad}^*_h\lambda \in\frak g^*$.
We take $\{e^i\}$, $\{e_i\}$ as dual bases of $\frak g^*$, $\frak g$ and $\{t^j\}$, $\{t_j\}$ dual bases of $\frak t^*$ and $\frak t$. A straightforward calculation gives that at each point $(g(x)h,\lambda)\in G\times \frak t^*_{\rm reg}$
\begin{eqnarray*}
{F_g}_*(l_h(e_i))&=&l_{gh}(e_i)+l_{gh}(h^{-1}g^{-1}\frac{\partial g}{\partial X^i}h),\\
{F_g}_*(\frac{\partial}{\partial t_j})&=&\frac{\partial}{\partial t_j}+l_{gh}(h^{-1}g^{-1}\frac{\partial g}{\partial T^j}h)
\end{eqnarray*}
where $X^i:=[{\rm Ad}_he_i,x]$, $T^j:={\rm Ad}^*_ht^j$ are tangent vectors at $x={\rm Ad}^*_h\lambda$. Note that $T^j\in\frak g_x$ (the isotropic subalgebra at $x$) and $X^i$ span the tangent space $T_x\frak g^*$ and thus the above formulas involve all the possible derivatives of $g\in {\rm Map}(\frak g^*,G)$.
A direct computation shows that at each point $(g(x)h,\lambda)\in G\times\frak t'$ (here $x={\rm Ad}^*_h\lambda\in\frak g^*$)
\begin{eqnarray*}
{F_g}_*(\pi)(g(x)h,\lambda)=\pi(g(x)h,\lambda)+l_{gh}(\otimes^2{\rm Ad}_{h^{-1}}U(x)),
\end{eqnarray*}
where $U(x)\in \frak g\wedge \frak g$ is defined, by using the notation in Theorem \ref{EEM}, as
\begin{eqnarray*}
U(x)=g_1^{-1}d_2(g_1)-g_2^{-1}d_1(g_2)+\langle {\rm id}\otimes {\rm id}\otimes x,[g_1^{-1}d_3(g_1),g_2^{-1}d_3(g_2)]\rangle.
\end{eqnarray*}
Thus by comparing with the expression of $\pi_r$,
\begin{eqnarray*}
\pi_r(g(x)h,\lambda)=\pi(g(x)h,\lambda)+l_{gh}(r_{\scriptscriptstyle \rm AM}(\lambda)-\otimes^2{\rm Ad}_{(gh)^{-1}}r_0),
\end{eqnarray*}
we obtain that ${F_g}_*(\pi)=\pi_r$ at point $(g(x)h,\lambda)\in G\times \frak t'$ if and only if
\begin{eqnarray*}
r_{\scriptscriptstyle \rm AM}(\lambda)=\otimes^2{\rm Ad}_{(gh)^{-1}}r_0+\otimes^2{\rm Ad}_{h^{-1}}U(x).
\end{eqnarray*}
Note that $x={\rm Ad}^*_h\lambda$, by the equivariance of $r_{\scriptscriptstyle \rm AM}$, we have $\otimes^2{\rm Ad}_hr_{\scriptscriptstyle \rm AM}(\lambda)=r_{\scriptscriptstyle \rm AM}(x)$. Thus the above formula is exactly the gauge transformation equation $r^g_0=r_{\scriptscriptstyle \rm AM}$.
\qed

\Addresses


\begin{thebibliography}{}


\bibitem{Anton}
A. Alekseev, {\em On Poisson actions of compact Lie groups on symplectic manifolds}, J. Differential Geometry 45 (1997), 241-56.

\bibitem{Anton1}
A. Alekseev and E. Meinrenken, {\em The non-commutative Weil algebra}, Invent. Math. 139 (2000), 135-72.

\bibitem{AMM}
A. Alekseev, A. Malkin, and E. Meinrenken, {\em Lie group valued moment maps}, J. Differential Geom, 48 (1998) 445–495.

\bibitem{AM}
A. Alekseev and A. Malkin, {\em Symplectic structures associated to Lie-Poisson groups}, Comm. Math. Phys. 162 (1994), 147 – 173.

\bibitem{AtiyahBott}
M. Atiyah and R. Bott, {\em The Yang-Mills equations over Riemann surfaces}, Philosophical Transactions of the Royal Society of London, Series A, Mathematical and Physical Sciences, 54 (1983) 523-615.

\bibitem{BBRS}
W. Balser, B.J.L. Braaksma, J.P. Ramis, and Y. Sibuya, {\em Multisummability of formal power series
solutions of linear ordinary differential equations}, Asymptotic Analysis 2 (1991), 27–45.

\bibitem{Feher}
J. Balog, L.Feh${\rm \acute{e}}$r and L. Palla, {\em Chiral extensions of the WZNW phase space, Poisson-Lie symmetries and
groupoids}, Nucl. Phys. B 568 (2000), 503-42.

\bibitem{BJL}
W. Balser, W.B. Jurkat, and D.A. Lutz, {\em Birkhoff invariants and Stokes' multipliers for meromorphic linear differential equations}, J. Math. Anal. Appl. 71 (1979), 48-94.

\bibitem{Boalch1} P. Boalch, {\em Stokes matrices, Poisson Lie groups and Frobenius manifolds}, Invent. Math. 146 (2001), 479–506.
no. 3, 479-506.

\bibitem{Boalch2} P. Boalch, {\em Symplectic manifolds and isomonodromic deformations}, Adv. in Math. 163 (2001), 137–205.

\bibitem{BoalchG} P. Boalch, {\em G-bundles, isomonodromy and quantum Weyl groups}, Int. Math. Res. Not. (2002), no. 22, 1129–1166.

\bibitem{Boalch3} P. Boalch, {\em Quasi-Hamiltonian geometry of meromorphic connections}, Duke Math. J. 139 (2007), no. 2, 369–405.

\bibitem{BTL1}
T. Bridgeland and V. Toledano Laredo, {\em Stokes factors and multilogarithms,} J. Reine Angew. Math. 682 (2013), 89–128.

\bibitem{Drinfeld} V. Drinfeld, {\em Quasi-Hopf algebras}, Leningrad Math. J. 1 (1990), no. 6, 1419-57.

\bibitem{Duistermaat}
J.J. Duistermaat, {\em On the similarity between the Iwasawa projection and the diagonal part}, Mem.
Soc. Math. France (1984), no. 15, 129–138.

\bibitem{EE}
B. Enriquez and P. Etingof, {\em Quantization of Alekseev-Meinrenken dynamical $r$-matrices}, AMS Transl. 210 (2003), no. 2, 81-98.

\bibitem{EEM}
B. Enriquez, P. Etingof and I. Marshall, {\em Comparison of Poisson Structures and Poisson-Lie dynamical r-matrices}, Int. Math. Res. Not. 2005, no. 36, 2183–2198.

\bibitem{EH}
B. Enriquez and G. Halbout, {\em Poisson algebras associated to quasi-Hopf algebras}, Adv. Math. 186 (2004), no. 2, 363-95.

\bibitem{EK}
P. Etingof and D. Kazhdan, {\em Quantization of Lie bialgebras, \uppercase\expandafter{\romannumeral1}, \uppercase\expandafter{\romannumeral2}}, Selecta Math. (N.S.) 2 (1996), no. 1, 1-41; 4 (1998), no. 2, 213-31.

\bibitem{EN}
P. Etingof and D. Nikshych, {\em Vertex-IRF transformation and quantization of dynamical $r$-matrices}, Math. Res. Lett. 8 (2001), no. 3, 331-45.

\bibitem{EV}
P. Etingof and A. Varchenko, {\em Geometry and classification of solutions of the classical
dynamical Yang-Baxter equation}, Commun. Math. Phys. 192 (1998) 77-129.

\bibitem{FM}
L. Feh${\rm \acute{e}}$r and I. Marshall, {\em The non-abelian momentum map for Poisson-Lie symmetries on the chiral
WZNW phase space}, Int. Math. Res. Not. 49 (2004), 2611-36.

\bibitem{Felder}
G. Felder, {\em Conformal field theory and integrable systems associated to elliptic curves}, Proc. ICM Z$\ddot{u}$rich, (1994), 1247-1255.

\bibitem{GW}
V. Ginzburg and A. Weinstein, {\em Lie-Poisson structure on some Poisson-Lie groups}, J. Amer. Math.
Soc. 5:2 (1992), 445-53.

\bibitem{GS}
V. Guillemin and S. Sternberg, {\em Symplectic techniques in physics}, C.U.P., Cambridge, 1984.

\bibitem{Krichever}
I. Krichever, {\em Isomonodromy equations on algebraic curves, canonical transformations and Whitham equations,} Mosc. Math. J., 2(4): 717–752, 2002.

\bibitem{Loday}
M. Loday-Richaud, {\em Stokes phenomenon, multisummability and differential Galois groups}, Ann. Inst.
Fourier 44 (1994), no. 3, 849–906.

\bibitem{Lu1}
J.-H. Lu and A. Weinstein, {\em Poisson Lie groups, dressing transformations, and Bruhat decompositions}, Journal of Differential Geometry 31 (1990), 501 - 526.

\bibitem{Lu}
J.-H. Lu and A. Weinstein, {\em Groupo${\rm \ddot{\iota}}$des symplectiques doubles des groupes de Lie-Poisson}, C. R. Acad. Sc. Paris, 309 (1989), 951 - 954.

\bibitem{MR}
B. Malgrange and J.-P. Ramis, {\em Fonctions multisommables}, Ann. Inst. Fourier (Grenoble) 42 (1992),
no. 1-2, 353–368.

\bibitem{STS}
M. Semenov-Tian-Shansky, {\em Dressing transformations and Poisson group actions}, Publ. Res. Inst. Math. Sci. 21 (1985), no. 6, 1237-1260.

\bibitem{TL}
V. Toledano Laredo, {\em Quasi-Coxeter quasitriangular quasibialgebras and the Casimir connection}, arxiv:1601.04076.

\bibitem{TLXu}
V. Toledano Laredo and X.-M. Xu, {\em Stokes phenomenon, Poisson Lie groups and quantum groups.} In preparation.

\bibitem{Wasow}
W. Wasow, {\em Asymptotic expansions for ordinary differential equations}, Wiley Interscience, New York,
1976.

\bibitem{Alan}
A. Weinstein, {\em Coisotropic calculus and Poisson groupoids}, J. Math. Soc. Japan. 40 (1988), 705–727.

\end{thebibliography}
\end{document}